\documentclass[article,prb,twocolumn,superscriptaddress,nobibnotes,showpacs]{revtex4-1}

\usepackage{graphicx}
\usepackage{amssymb}
\usepackage{bm}
\usepackage{tabularx}
\usepackage[normalem]{ulem}

\usepackage{amsmath}
\usepackage[dvipsnames]{xcolor}

\usepackage[colorlinks=true,
  		linkcolor=black,
	        citecolor=blue,
	        urlcolor=blue]{hyperref}

\newcommand{\marvel}{National Center for Computational Design and Discovery of Novel Materials~(MARVEL), \'Ecole Polytechnique F\'ed\'erale de Lausanne, CH-1015 Lausanne, Switzerland}
\newcommand{\dqmp}{Department of Quantum Matter Physics, University of Geneva, 24 Quai Ernest Ansermet, CH-1211 Geneva, Switzerland}
\newcommand{\gap}{Group of Applied Physics, University of Geneva, 24 Quai Ernest Ansermet, CH-1211 Geneva, Switzerland}	

\makeatletter
\g@addto@macro\bfseries{\boldmath}
\makeatother
\AtBeginDocument{\usepackage{booktabs}}
\begin{document}
\title{Multi-frequency Shubnikov-de Haas oscillations in topological semimetal Pt$_2$HgSe$_3$}
\date{\today}
\author{Diego Mauro}
\affiliation{\dqmp}
\affiliation{\gap}
\author{Hugo Henck}
\affiliation{\dqmp}
\affiliation{\gap}
\author{Marco Gibertini}
\affiliation{\dqmp}
%\affiliation{\gap}
\affiliation{\marvel}
\author{Michele Filippone}
\affiliation{\dqmp}
%\affiliation{\gap}
\author{Enrico Giannini}
\affiliation{\dqmp}
%\affiliation{\gap}
\author{Ignacio Guti\'errez-Lezama}
\affiliation{\dqmp}
\affiliation{\gap}
\author{Alberto F. Morpurgo}
\affiliation{\dqmp}
\affiliation{\gap}

\begin{abstract}
Monolayer jacutingaite (Pt$_2$HgSe$_3$) has been recently identified as a candidate quantum spin Hall system with a 0.5 eV band gap, but no transport measurements have been performed so far on this material, neither in monolayer nor in the bulk. By using a dedicated high-pressure  technique, we grow crystals enabling the exfoliation of 50-100 nm thick layers and the realization of devices for controlled transport experiments. Magnetoresistance measurements indicate that jacutingaite is a semimetal, exhibiting Shubnikov-de Haas (SdH) resistance oscillations with a multi-frequency spectrum. We adapt the Lifshitz-Kosevich formula to analyze quantitatively the SdH resistance oscillations in the presence of multiple frequencies, and find that the experimental observations are overall reproduced well by band structure ab-initio calculations for bulk jacutingaite. Together with the relatively high electron mobility extracted from the experiments ($\approx 2000$ cm$^2$/Vs, comparable to what is observed in WTe$_2$ crystals of the same thickness), our results indicate that  monolayer jacutingaite should provide an excellent platform to investigate transport in 2D quantum spin Hall systems.

\end{abstract}

\maketitle

\begin{figure*}%%[ht]
  \includegraphics[width= 0.9\textwidth]{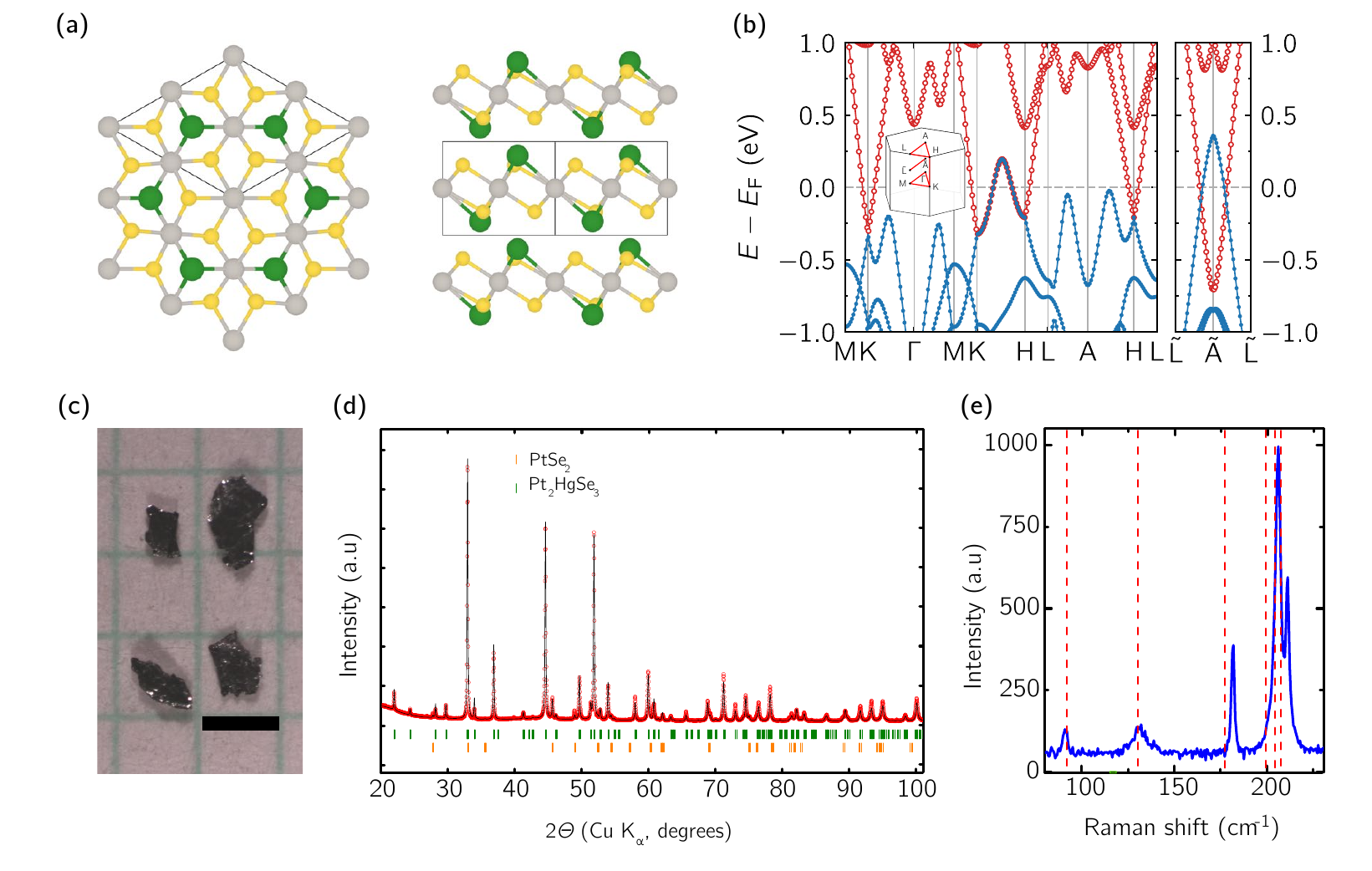}
  \caption{ (a) Top and side view of the jacutingaite ($\mathrm{Pt_2HgSe_3}$) structure, with the green, grey, and yellow balls representing  mercury, platinum, and selenium atoms, respectively.  Thin black solid lines represent the edges of the primitive unit cell. (b) Calculated band structure for bulk jacutingaite in the absence of spin-orbit interaction; valence (conduction) bands are represented with blue (red empty) circles. The degeneracy along the $\mathrm{K}\mathrm{H}$ line, responsible for the presence of a nodal line, is clearly visible. Band inversion occurs along the $\mathrm{\tilde L}\mathrm{\tilde A}\mathrm{\tilde L}$ line. (c) Optical microscope image of jacutingaite crystals obtained from the growth process that we have developed (the bar is 1 mm). (d) X-ray powder diffraction pattern from a ground piece of the as-grown material. Red symbols represent the experimental data and the continuous black line the result of the Rietveld structural refinement that considers the presence of both $\mathrm{Pt_2HgSe_3}$ and $\mathrm{PtSe_2}$  (green and orange markers indicate the position of diffraction lines of $\mathrm{Pt_2HgSe_3}$ and $\mathrm{PtSe_2}$, respectively). (e) Raman spectra measured on as-grown jacutingaite crystals. Several Raman active modes are present at wavelength close to those predicted by ab-initio calculations (marked by the red dashed lines).}
  \label{fig:Fig1}
\end{figure*}

 %AFM image and corresponding profile (red continuous line) of a jacuitngaite layer exfoliated onto $\mathrm{Si/SiO_2}$ (the bar is $2 \mathrm{ \mu m}$)

\section{Introduction}

Two-dimensional (2D) quantum spin Hall insulators (QSHIs) constitute an important  class of topological systems having  a gapped insulating bulk and gap-less helical edge states that mediate transport at low energy~\cite{kane_quantum_2005,kane_z2_05,Bernevig2006,Konig2007,Knez2011,Suzuki2013,Fei2017,Wu2018}. According to theory, these helical states are protected against elastic backscattering by time-reversal symmetry, so that edge conduction is predicted to occur in the ballistic regime over distances shorter than the inelastic scattering length. In reality, the material systems that have been identified experimentally as 2D QSHI --such as HgTe/CdTe~\cite{Konig2007, Knez2011,Suzuki2013,Roth2009,Grabecki2013,Konig2013,Wu2018} and InAs/GaSb~\cite{Konig2007, Knez2011,Suzuki2013,Roth2009,Grabecki2013,Konig2013,Wu2018} quantum wells-- do not exhibit the expected, precisely quantized conductance: values close to $e^2/h$ per edge have been reported, with rather large sample-to-sample variations, as well as fluctuations as a function of applied gate voltage. Identifying the physical origin of these deviations is proving difficult, although several mechanisms have been put forward\cite{Konig2007,Wu2006,Schmidt2011,Budich2012,Vayrynen2013,Wang2017,Novelli2019}. To progress, it is important to investigate experimentally a broader variety of 2D QSHI systems, in order to understand the influence of different microscopic parameters.

2D van der Waals (vdW) layered materials offer new opportunities, as illustrated by $\mathrm{WTe_2}$ monolayers, whose predicted\cite{Qian2014} 2D QSHI nature has been confirmed by different experiments\cite{Fei2017,Tang2017,Wu2018,Shi2019, Cucchi2019}. Two-terminal edge conductance close to $e^2/h$ has been observed in hBN-encapsulated $\mathrm{WTe_2}$ monolayers\cite{Fei2017}, with subsequent work reporting a more precise quantization (albeit only for contact separation shorter than 100 nm \cite{Wu2018}). Edge states have been directly detected by means of scanning tunneling microscopy \cite{Tang2017} and other types of local probes \cite{Shi2019}, and even the bands of $\mathrm{WTe_2}$ monolayers could be mapped by means of angle-resolved photo-emission experiments (ARPES)\cite{Tang2017,Cucchi2019}. The 60-70 meV band-gap --significantly larger than in previous 2D QSHI systems-- and the band structure were found to be in  agreement with ab-initio calculations\cite{Qian2014,Tang2017,Cucchi2019,Zheng2016}. The improved conductance quantization, the larger band-gap, as well as the possibility to access directly the material with different experimental probes make $\mathrm{WTe_2}$ an excellent candidate to study the physics of 2D QSHIs. These results also showcase the potential of 2D materials, and motivate the search for other atomically thin crystals\cite{dresden_htz2_2018,olsen_z2ht_2019,Marrazzo2019}, providing even better platforms for experimental work.

Among these, jacutingaite --a vdW layered material with chemical formula $\mathrm{Pt_2HgSe_3}$-- has been proposed to be of particular interest \cite{Marrazzo2018}. In its bulk form (see the crystal structure in \autoref{fig:Fig1}a), jacutingaite is predicted  to be a compensated semimetal hosting multiple electron and hole pockets that arise from nodal lines along which the valence (blue) and conduction (red) bands are degenerate in the absence of spin-orbit interaction (SOI) (see e.g.\ the KH line in \autoref{fig:Fig1}b). SOI is, however, not negligible owing to the presence of heavy elements and lifts the degeneracy between valence and conduction bands (that remain separately two-fold degenerate owing to the combination of inversion and time-reversal symmetry), endowing the system with a non-trivial topology protected by crystalline symmetries\cite{facio_prm_2019,bansil_arxiv_2019,theory_jacu_2019}. The correspondingly expected surface states have been recently observed in ARPES measurements\cite{exp_jacu_2019}, and found to be in remarkable good agreement with theory\cite{facio_prm_2019,bansil_arxiv_2019,theory_jacu_2019}. When the material thickness is reduced to that of an individual monolayer, jacutingaite is predicted to become a 2D QSHI realizing the original Kane-Mele model\cite{kane_quantum_2005,kane_z2_05}, with a $\sim$ 0.5 eV band gap~\cite{Marrazzo2018} (one order of magnitude larger than in $\mathrm{WTe_2}$). Calculations also suggest that jacutingaite monolayers can be switched between a topological and a trivial insulating state, by applying electric fields \cite{Marrazzo2018}, as possibly achievable in 2D vdW heterostructures. If confirmed, all these predictions make monolayer jacutingaite an ideal platform to explore --and possibly control-- 2D QSHI systems.

Despite a great interest in jacutingaite\cite{Marrazzo2018,theory_jacu_2019,Cucchi2019,Kandrai2019,facio_prm_2019,bansil_arxiv_2019,Wu2019}, no transport measurements have been performed so far, neither on atomically thin layers nor on bulk, largely because growing high-quality crystals of sufficiently large size is challenging\cite{cabral_first_obs_08}. Here, we solve this problem by employing a high-pressure crystal synthesis technique to grow millimiter-sized crystals, which we exfoliate into thin layers (typically 50-to-100 nm thick) that enable the realization of multi-terminal devices for controlled transport experiments, and allow the quality of nano-fabricated structures to be assessed. The measured magnetoresistance indicates the presence of both electrons and holes, and exhibits pronounced Shubnikov-de Haas (SdH) oscillations with multiple frequencies. We discuss in detail how the Lifshitz-Kosevich (LK) formula can be adapted to the case of jacutingaite, to extract quantitative information. We find that the SdH oscillations are due to electrons, and that the observed frequencies originate from distinct pockets in the conduction band in overall good agreement with first-principles calculations. These results demonstrate that high-quality devices based on exfoliated layers can be produced (at least as good as WTe$_2$ crystals of the same thickness\cite{Wang2015,Wang2016}), a necessary step towards the investigation of the 2D QSHI state expected in monolayers that confirms the experimental potential of jacutingate.

\begin{figure*}%%[ht]
  \includegraphics[width= 0.75\textwidth]{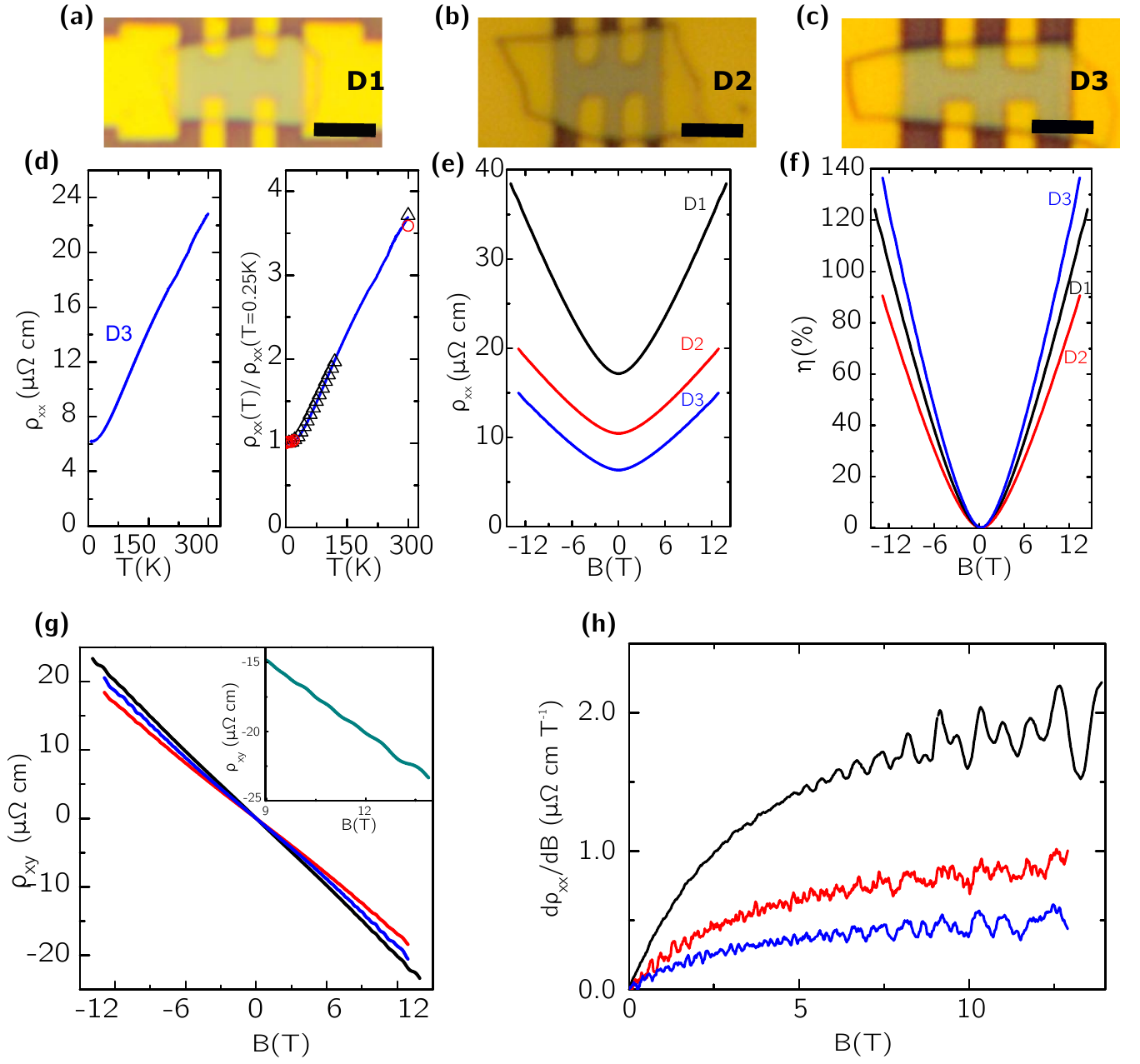}
  \caption{ (a), (b), and (c) Optical microscope images of device D1, D2 and D3 (the scale bar in each panel is $2 \mu$m). (d) Left: Temperature dependence of the four-probe resistivity measured on D3 while the device was slowly cooled from room temperature to 0.25 K. Right: temperature dependence of the resistivity normalized to the value measured at $T=0.25$ K, $\rho_{xx}(T)/\rho_{xx}(T = 0.25 K) $ for device D1 (black triangles), D2 (red circles), and D3 (blue continuous line); for D1 and D2 the resistivity was measured at room-temperature and at the different temperatures at which Shubnikov-de Haas oscillations have been observed, whereas for D3 the resistivity was measured as a function of $T$ while cooling down the device slowly from room temperature. (e)  Longitudinal resistivity ($\rho_{xx}$) as a function of magnetic field $B$ measured on device D1, D2, and D3 at $T= 250~\mathrm{mK} $, and (f) corresponding magnetoresistance $\eta \equiv (\rho_{xx}(B)-\rho_{xx}(0))/\rho_{xx}(0) $. (g)  Transverse resistivity ($\rho_{xy}(B)$) for the three devices investigated at 250~mK, showing a negative slope indicative of electron-type transport. Inset: zoom-in on the high-field transverse magnetoresistance of device D1, showing the presence of SdH oscillations. (h) Derivative of the longitudinal resistivity $\rho_{xx}(B)$ that puts in evidence the SdH oscillations, which start from 5-7 T, depending on the device. In all panels black, red and blue curves represent data measured on D1, D2, and D3.  }
  \label{fig:Fig2}
\end{figure*}

\section{Methods}
The small size ($\leq 50 \mathrm{\mu m}$) of existing jacutingaite crystals --extracted either from aggregates of platinum based mined minerals \cite{cabral_first_obs_08}, or from polycristalline powders grown in the laboratory \cite{jacutingaite_exp_12}-- has so far prevented the exfoliation of thin layers to realize multi-terminal devices for controlled transport experiments. To grow larger high-quality crystals we have employed a high-pressure synthesis process that allows the growth of suitable materials in a few hours (see Supporting Information, SI, for details). The process results in a conglomerate a few millimeters in size, containing single crystals (see \autoref{fig:Fig1}c) with dimensions up to  $\sim 0.6 \times 1 \, \mathrm{mm^2}$. X-ray powder diffraction (XRD) shows the presence of crystals of two distinct compounds in the conglomerate, which we successfully identify as  $\mathrm{Pt_2HgSe_3}$ and $\mathrm{PtSe_2}$ from the corresponding series of diffraction peaks in \autoref{fig:Fig1}d (the green and orange markers indicate the diffraction peaks for jacutingaite and $\mathrm{PtSe_2}$; the majority of crystals are jacutingaite, and 6-7 \% of the material consists of PtSe$_2$). To ensure that the material used for exfoliation is actually jacutingaite, we performed wavelength-dispersive X-ray spectroscopy (WDS)  on each individual crystal and determine its chemical composition. We find the stoichiometry of the jacutingaite crystals to correspond to the nominal one within the sensitivity of our WDS system (approximately 1\%), uniformly across the individual crystals. The identification of the jacutingaite crystals is confirmed by Raman spectroscopy measurements, which exhibit a series of active modes at characteristic wavelengths, in good agreement with values calculated for jacutingaite (see \autoref{fig:Fig1}e red dashed lines and SI).

For the device fabrication we exfoliate thin jacutingaite layers and transfer them onto $\mathrm{Si}$ substrates covered with 300 nm $\mathrm{SiO_2}$. Conventional exfoliation procedures based on adhesive tape result in layers with lateral dimensions up to $\sim 10\mathrm{\mu m}$ and thickness in the  50-100 $\mathrm{nm}$ range. Thinner crystals can also be found (down to few nanometers), but their lateral dimension  ($\sim 1 \mathrm{\mu m}$) is normally too small to fabricate multi-terminal devices (as it is the case for virtually all other 2D materials, the availability of larger crystals --for which we are currently modifying the growth process-- will likely yield larger exfoliated thin layers and eliminate this problem). Prior to contact deposition, we perform again Raman spectroscopy measurements, to ensure that the selected exfoliated layer is not an inclusion of  $\mathrm{PtSe_2}$, whose known Raman spectrum \cite{O_Brien_2016} is distinctly different from that of jacutingaite (see \autoref{fig:Fig1}e). We found in all cases that exfoliation of jacutingaite crystals never shows any sign of PtSe$_2$ inclusions, indicating that during the growth the two materials phase separate, i.e. the crystals in the as grown conglomerate are either jacutingaite or PtSe$_2$.  Contacts are prepared in a Hall-bar geometry (see top panel of \autoref{fig:Fig2}a), using conventional nano-fabrication techniques (i.e., e-beam lithography, Ti/Au evaporation, and lift-off). As jacutingaite is a natural occurring mineral, the material is expected to be stable under ambient conditions, and indeed even the thinnest layers that we have exfoliated do not exhibit any  visual sign of degradation, neither after several months of exposure to air nor after exposure to the chemicals commonly used in the device fabrication processes. For this reason, all the individual fabrication steps have been performed in air.

\section{Results and discussion}
We measured three different devices (see figure \autoref{fig:Fig2}a; labeled as D1, D2 and D3), based on layers that are respectively 90, 50 and 90 $\mathrm{nm}$ thick, all exhibiting similar behavior. The temperature dependence of the longitudinal resistivity $\rho_{xx}$ of device D3, measured while slowly cooling the device from room temperature down to 250 mK, is shown in \autoref{fig:Fig2}b (left panel). As expected for a semimetal, metallic behavior is observed, with $\rho_{xx}$ decreasing linearly from 23$ \; \mathrm{\mu \Omega \cdot cm} $ at room temperature to 6.2 $\; \mathrm{\mu \Omega \cdot cm}$ at 15~K, below which saturation occurs similarly to what is observed for the resistivity of WTe$_2$ crystals of comparable thickness \cite{Wang2015}. Finding that $\rho_{xx}$ decreases upon lowering temperature indicates that down to $\sim 15 $~K transport is limited by inelastic scattering and not by collisions with impurities, and provides a first indication of the electronic quality of our crystals. A very similar temperature dependence of the resistivity is observed for all the three devices measured, as can be seen in the right panel of \autoref{fig:Fig2}b, which shows the resistivity normalized to its value at 250 mK $\rho_{xx}(T)/ \rho_{xx}({T = 0.25 K})$. Data measured on the three devices (black, red and blue symbols for D1, D2 and D3 respectively) collapse onto a single curve, indicating that our exfoliated jacutingaite crystals are of similar quality, and exhibit a virtually identical residual resistivity ratio $RRR = \rho (T=300K)/ \rho (T=0.25K)$ (3.71, 3.59, 3.69 for device D1, D2 and D3, respectively). Unexpectedly, in contrast with this behavior, the absolute value of the resistivity appears to exhibit significant differences in the three measured devices, as it may be inferred from the resistivity values measured at 250 mK --respectively $\sim 17.1 \;$, $10.4 \;$ and $ 6.3~\mathrm{\mu \Omega \cdot cm}$ in D1, D2, and D3 (see $B=0$ values in \autoref{fig:Fig2}c). This seemingly large spread is mainly a consequence of the difficulty in determining the correct length-to-width ratio in our devices, which is affected by a sizable uncertainly because of the small dimensions of the exfoliated crystals  (see \autoref{fig:Fig1}a; indeed, the $\rho_{xx}(T)/ \rho_{xx}(T = 0.25 K)$ ratio does not depend on the device dimensions, which is why an excellent reproducibility is observed when looking at this quantity). Up to a smaller extent, the observed sample-to-sample deviations in absolute value of resistivity are also due to differences in the carrier mobility, as we discuss in more detail below.

The application of a perpendicular magnetic field up to 14 T, the highest field reached in our experiments (see \autoref{fig:Fig2}c), causes the resistivity to increase. The corresponding magneto-resistance (MR; $\eta \equiv (\rho_{xx}(B)-\rho_{xx}(0))/\rho_{xx}(0) $), shown in figure \autoref{fig:Fig2}d, reaches values ranging from 80 \% to 140\%, depending on the device. Notably, at high magnetic fields the MR deviates from the quadratic behavior commonly expected from classical dynamics of simple conductors, as $\eta$ depends linearly on $B$. A linear MR has been reported previously in systems with topological properties and/or strong SOI\cite{Xu2011,Liang2015,Shekhar2015,Leahy10570}, but the precise mechanism at work in jacutingaite remains to be determined.

The magnetic field dependence of the transverse resistivity $\rho_{xy}$ is shown in \autoref{fig:Fig2}e. With the exception of device D3 (that exhibits a small, but clear, non-linearity), the $\rho_{xy}(B)$ curve is essentially linear, with a negative slope indicative of a dominating electron contribution to transport. If interpreted in terms of a single type of charge carriers we find a spread of about 20\% in the density values extracted from different devices: $n = 3.75 \cdot 10^{20} \, \mathrm{cm}^{-3}$ , $ 4.08 \cdot 10^{20} \, \mathrm{cm}^{-3}$ and $4.51 \cdot 10^{20}\, \mathrm{cm}^{-3}$ for D1, D2 and D3. Such a spread (as well as the fact that these values are twice laerger than what is expected from ab-initio calculations, $n = 2 \cdot 10^{20} \, \mathrm{cm}^{-3}$) cannot be attributed to unintentional doping due to impurities, since this explanation would require impurity concentrations reaching up to $\sim$ 5\%. Such a concentration would not only have been detected in the WDS characterization of the material (see above), but would also cause a very short electronic mean free path, incompatible with the presence of the well-defined SdH conductance oscillations discussed below.

Finding inconsistencies when attempting to interpret the Hall data in terms of a single type of charge carrier is not surprising, because jacutingaite is predicted to be a compensated semimetal with an equal density of electrons and holes. Under the simplifying assumption that only one family of electrons and one family of holes are present, the expression for the Hall slope becomes $R_H =-\frac{1}{ent}\frac{1-(\mu_h/\mu_e)}{1+(\mu_h/\mu_e) }$ \cite{ziman1972principles} ($\mu_h$ and $\mu_e$ are the hole and electron mobility). This expression naturally explains the larger densities extracted above, because --if the Hall slope is interpreted in terms of only electrons-- a larger value of $n$ is obtained as compared to the actual one since $\frac{1-(\mu_h/\mu_e)}{1+(\mu_h/\mu_e)}<1$. To account for the observed difference in Hall slopes it suffices that the $\mu_h/\mu_e$ ratio varies by a factor of 0.5 --i.e. that the hole mobility is approximately two times smaller than the electron mobility in different devices-- as it can certainly be expected (graphene exfoliated on SiO$_2$, for which the spread in carrier mobility is at least as large, can be taken as a term of comparison). In practice, however, the simultaneous presence of electrons and holes prevents the unambiguous and precise determination of the carrier density and mobility from Hall resistance data, because a reliable quantitative analysis needs to take into account the presence of multiple pockets for both types of charge carriers (each with a possibly different mobility value).

\begin{figure*}%[ht]
    \centering
    \includegraphics[width= 1 \textwidth]{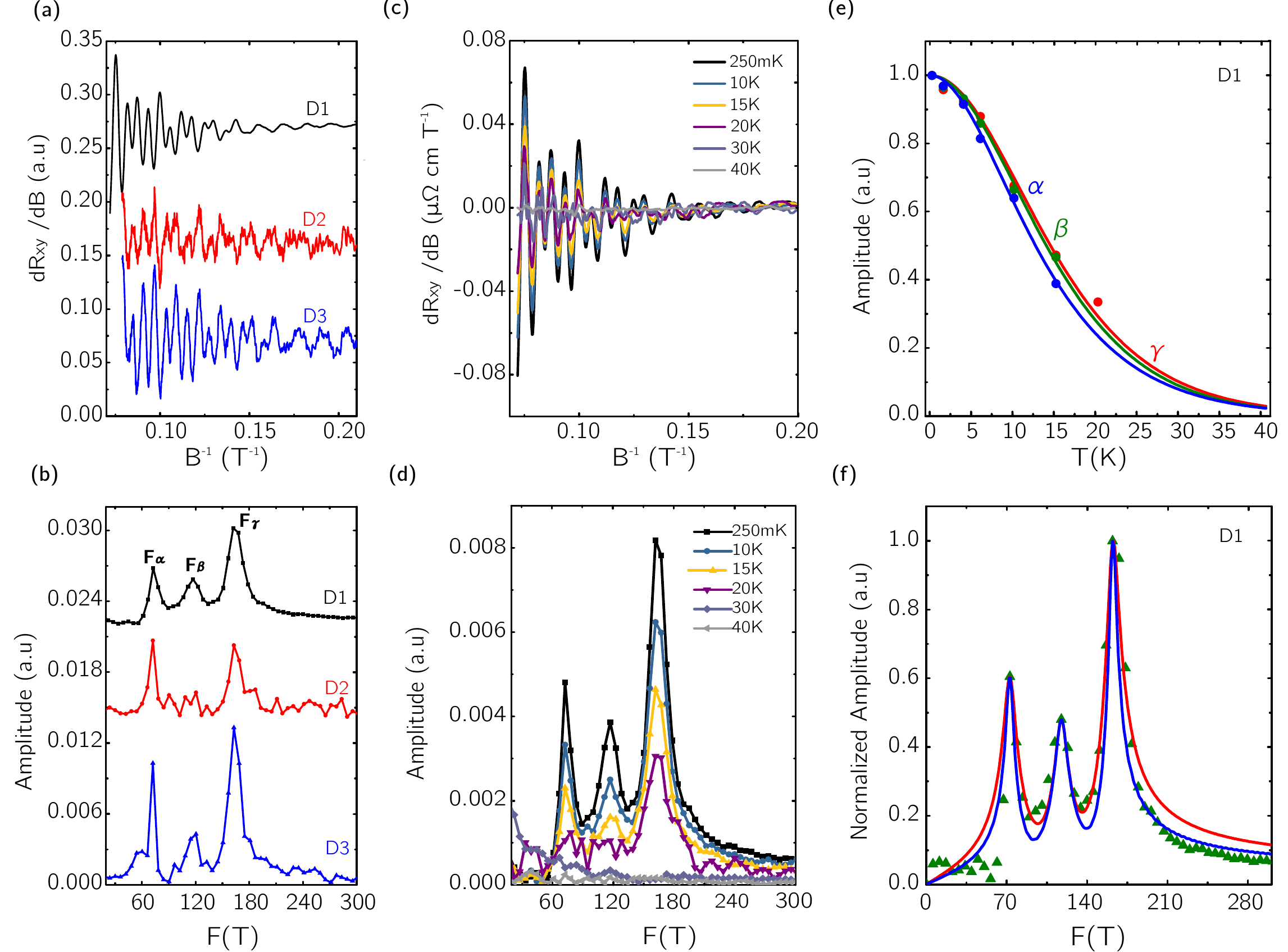}
    \caption{ (a) Derivative of the transverse resistivity $d\rho_{xy}/dB$ measured at 250 mK plotted as a function of $1/B$ (offset for clarity), exhibiting Shubnikov de Haas oscillations, and (b) corresponding Fourier Transform (black, red and blue curves represent data taken on device D1, D2, D3). Three peaks are present, located at the same frequencies  $F_{\alpha}=73$ T, $F_{\beta} = 118 $ T and $F_{\gamma} = 163 $ in the three devices (in D2, the peak at $F_{\beta}$ has virtually vanishing amplitude). (c) Temperature dependence of the SdH oscillations as extracted from the $d\rho_{xy}/dB$ data measured on device D1, and (d) corresponding Fourier transforms (in (c) and (d) the curves have been measured at the temperature values indicated in the legends).  (e) The color dots represent the normalized amplitude of the peaks at frequency $F_{\alpha}$ (blue) $F_{\beta}$ (green), and $F_{\gamma}$ (red) as a function of temperature, measured on device D1. The continuous lines of the corresponding color represent the best fit obtained from the FT of the LK formula as described in the main text, using the cyclotron mass (different for every peak) as only fitting parameter. (f) Comparison of the full functional dependence of the FT of $d\rho_{xy}/dB$ data measured on device D1 at 250 mK (green triangles) with the FT calculated from the LK formula. The red curve corresponds to the case in which the carrier mobility is fixed at the value determined by the experimentally observed onset of the SdH oscillations; in the blue curve, the carrier mobility value was chosen to be different for each FT peak (see main text).  }
    \label{fig:Fig3}
\end{figure*}

More information can be gained by analyzing the SdH conductance oscillations present in all our devices, visible in both the $\rho_{xx}(B)$ and $\rho_{xy}(B)$ curves (see the inset of \autoref{fig:Fig2}d). To facilitate their analysis we look at the derivative of $\rho_{xx}$ and $\rho_{xy}$, shown in \autoref{fig:Fig2}e and \autoref{fig:Fig3}a. An estimate of the charge carrier mobility is obtained from the magnetic field $B_{onset}$ at which the oscillations start, using the criterion $\mu B_{onset} \sim 1$; $B_{onset} \simeq 5-7$ T depending on the specific device implies mobility values in the range $\mu_e \sim 1400 - 2000 \; \mathrm{cm^2/V\cdot s}$. The values correspond to the electron mobility, as the sign of the Hall slope indicates that it is the electrons that give the dominant contribution to transport.

When plotted against $1/B$ the oscillations exhibit an irregular envelope (\autoref{fig:Fig3}a), originating from the presence of  multiple frequencies. Indeed, three peaks are present in the amplitude of the Fourier transform (FT), shown in \autoref{fig:Fig3}b, centered at frequencies $F_{\alpha}=73$ T, $F_{\beta} = 118 $ T and $F_{\gamma} = 163 $ T. The frequencies coincide in the three different devices (albeit the amplitude of the peak at $F_{\beta}$ nearly vanishes in device D2), indicating that the areas of the extremal Fermi surfaces responsible for the SdH oscillations are the same in all cases. It directly follows that --within the material band structure-- the Fermi level is located at the same energy in the different devices, and therefore that also the charge carrier density is the same. This confirms our interpretation of the Hall data, namely that the sample-to-sample dependence of the Hall slope is due to variations in the $\mu_h/\mu_e$ ratio, with $\mu_h$ that is approximately two times smaller than $\mu_e$. Note that with such $\mu_h$ values the onset of hole SdH oscillations is expected to occur at approximately  $\sim 12-15 $ T in the best devices, so that quantum oscillations of holes are not visible in our experiments.

The evolution of the SdH oscillations upon increasing temperature (see Fig. 3(c) for data measured on device D1) is of particular interest as it allows the cyclotron mass to be determined. The analysis of the data relies on the Lifshitz-Kosevich expression for the $T$ and $B$ dependence of the oscillation amplitude $\Delta\rho$ \cite{Schoenberg,LK,Richards1973,Niederer1974,Fang1977}:
\begin{equation}
\frac{\Delta\rho}{\rho } \propto \sqrt{\frac{\hbar e B}{m_c E_F}}\frac{X}{\sinh{X}} \exp\left ( \frac{- \pi }{\mu B} \right ) \sin\left ( \frac{2 \pi F}{B} \right )
\label{eqn:eq1}
\end{equation}
where $X = (2 \pi^2k_B T m_c / \hbar e B)$ and $k_B$ is the Boltzmann constant, $T$ is the temperature. The presence of multiple SdH oscillation frequencies, however, makes the situation complex. That is why we start by discussing in detail how the LK expression can be used in the present context to extract reliable quantitative information.

The common way to extract the cyclotron mass $m_c$ from SdH oscillations consists in plotting a resistance minimum measured at a fixed value of $B$ as a function of $T$, and subsequently fitting the data to the $X/\sinh X$ dependence predicted by Eq. (1) . This procedure works well in the presence of a single oscillation frequency, but it does not work if multiple frequencies are present. Without going into the microscopic physical details, that is because in the presence of multiple frequencies, the magnetic field values at which the SdH oscillation minima occur are found to vary upon changing $T$ (contrary to the case in which a single SdH oscillation frequency is present). Hence, it is simply not possible to analyze the resistance minima at a fixed value of $B$ as a function of $T$.

The issue is well-known from past work on materials exhibiting SdH oscillations with multiple frequencies\cite{Balicas2000,Bangura2008,Cai2015,Rhodes2015}. In those systems, a commonly employed heuristic approach  to analyze the temperature dependence of SdH oscillations consists in plotting the amplitude of a FT peak as a function of $T$ and fitting it with the expression $X/\sinh X$, i.e., with the same functional dependence normally used to fit the $T$-dependence of the resistivity minima. Although believed to give reasonable estimates for the cyclotron masses, this approach is problematic because the temperature dependence of the FT peak, as calculated from the LK expression, is not proportional to $X/\sinh X$. A simple way to understand the problem is to realize that --since $X = (2 \pi^2k_B T m_c / \hbar e B)$-- fitting the temperature dependence with the expression $X/\sinh X$ requires a specific value of magnetic field $B$ to be selected. The FT peak amplitude, however, is calculated by integrating the SdH oscillations over a large interval of magnetic field and not at one specific value of $B$. Therefore, the choice for the value of $B$ to be inserted in $X$ is arbitrary, which prevents a truly reliable determination of $m_c$.

We adopt a different approach, free from any ambiguity, that eliminates these problems. It consists in calculating the FT of the LK expression (Eq. (1)) and comparing the $T$-dependence of the calculated FT peak amplitude to the $T$-dependence of the corresponding experimentally measured quantity, using the cyclotron mass as a fitting parameter. The approach is free from ambiguity, because it compares the same quantity, measured or calculated.

\autoref{fig:Fig3}d shows the FT of the SdH oscillations measured on device D1 for different values of $T$, ranging from 250 mK to 40 K. The colored dots in \autoref{fig:Fig3}e represent the heights of the three FT peaks (normalized to the height at the lowest temperature of our experiments, i.e. 250 mK), extracted from the data in \autoref{fig:Fig3}d. The continuous lines of the corresponding color are best fits obtained by using the FT of the LK formula, calculated in the same $B$-range used to calculate the FT of the experimental data \footnote{The experimental spectra used in the analyses are obtained by calculating the FT of the the derivative of the SdH oscillations, for consistency we use the FT spectra of the derivative of equation \eqref{eqn:eq1}}. For each peak, the only fitting parameter is the cyclotron mass, as for all three peaks the carrier mobility (that also enters the theoretical expression) is fixed to be the value determined by the onset of the SdH oscillations. The agreement between data and the curves calculated from the LK expression is excellent and the value of the cyclotron mass extracted for each peak is given in \autoref{tab:masses} (for the peak corresponding to the frequency $F_{\beta}$, the quality of the data only allows us to determine the mass for device D1).

{

\newlength\lengthb \setlength\lengthb{8mm}

\begin{table}%%[ht]
\caption{\label{tab:masses} Effective masses $m_{\alpha}$, $m_{\beta}$, $m_{\gamma}$ extracted from the analysis of the temperature dependence of the amplitude of the $\alpha$, $\beta$ and $\gamma$  peaks in the FT, obtained from SdH resistance oscillations measured on devices D1, D2 and D3. The peak corresponding  to  the  frequency F$_{\beta}$ is seen over a sufficiently large temperature range to enable the extraction of the corresponding cyclotron mass only in device D1. Note how the three different devices give nearly coinciding values for the cyclotron masses for the $\alpha$ and the $\gamma$ peaks.}
\medskip
\centering
\begin{tabular}{@{} l
                @{\hspace*{2mm}}     c
                @{\hspace*{\lengthb}}c
                @{\hspace*{\lengthb}}c
                @{\hspace*{\lengthb}}c
                 @{}}
\toprule
          Device & $m_{\alpha} / m_e$ & $m_{\beta} / m_e$ & $m_{\gamma} / m_e$  \\
\midrule
D$_{1}$ & 0.110 & 0.135 & 0.130  \\
\addlinespace[0.2cm]
D$_{2}$  & 0.110 &   -    & 0.127  \\
\addlinespace[0.2cm]
D$_{3}$ & 0.103 &   -    & 0.125  \\
\bottomrule
\end{tabular}

\end{table}
}

The procedure can be applied not only to the analysis of the $T$-dependence of the peak height, but also to the full functional dependence of the peaks, i.e. to their line-shape. This is useful because at sufficiently low temperature, the peak width is determined by the carrier mobility, and the comparison  between calculated and measured FT can enable a more precise determination of this quantity. We illustrate the concept for device D1, for which the blue curve  in \autoref{fig:Fig3}f is calculated assuming $\mu_{\alpha} \simeq  1800 \, \mathrm{cm^2/V\cdot s}$,  $\mu_{\beta} \simeq  1000 \, \mathrm{cm^2/V\cdot s}$ and  $\mu_{\gamma} \simeq 1800 \, \mathrm{cm^2/V\cdot s}$ for the mobility values of the electrons in the pockets responsible for the different peaks in the FT. It is apparent that with these values the agreement with the data at high frequency is improved.
Although a precise optimization is not particularly relevant in the present case, it is important to realize that using the LK formula properly to analyze the FT of the SdH oscillations offers possibilities that go beyond the heuristic approach that is commonly employed.

\begin{figure*}%%[ht]
    \centering
    \includegraphics{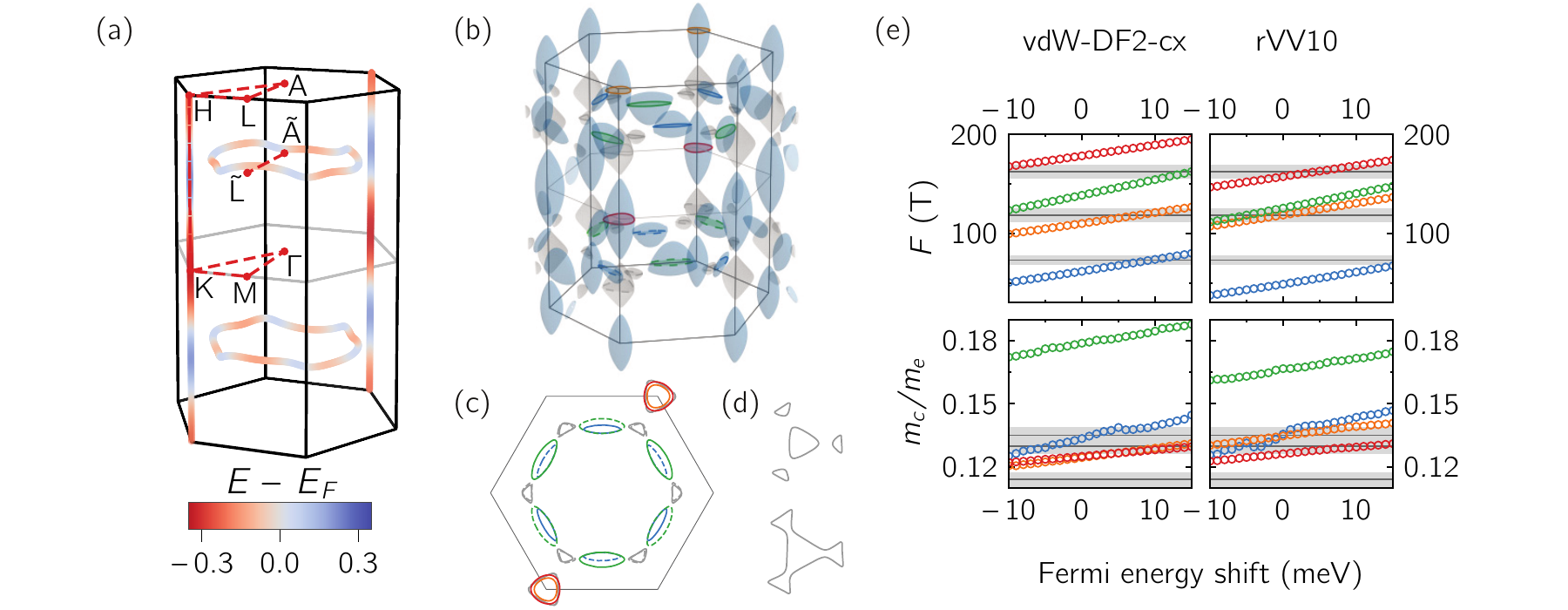}
    \caption{(a) The two sets of nodal lines in the Brillouin zone of jacutingaite: vertical nodal lines are present at the edge of the Brillouin zone (BZ) and other closed-loop nodal lines encircle the the $\Gamma$A direction. The color coding denotes the energy relative to the Fermi level. The red dashed lines show the paths over which energy bands \autoref{fig:Fig1}b are computed. (b) Three-dimensional representation of the calculated Fermi surface in reciprocal space (the thin solid line represents the boundary of the BZ). Blue (grey) surfaces denote electron (hole) pockets, whose extremal contours are highlighted with solid (dashed) lines of different colors for positive (negative) values of $k_z$. In agreement with panel (a), we find alternating electron and hole pockets along either the vertical nodal lines or horizontal nodal loops. (c) Projections on the $(k_x,k_y)$ plane of the extremal contours resulting from all different pockets, with the same colors and line styles as in panel (b). (d) Evolution of the hole pockets around the KH line, merging into a single one as seen in ARPES when the Fermi energy is slightly shifted on the hole side. (e) SdH oscillation frequencies $F$ (in Tesla) and cyclotron effective masses $m_c$ (in units of the free electron mass $m_e$) calculated for structures relaxed using two different functionals (vdW-DF2-cx\cite{Dion2004,Berland2014} and rVV10 \cite{Vydrov2010,Sabatini2013} ), as a function of the shift in Fermi energy with respect to the DFT neutral reference. The range of Fermi energy considered corresponds to density variations of the order of $1\times10^{20}$~cm$^{-3}$ (see SI). Different colors correspond to different electron pockets with the same coding as in panels (b) and (c). Experimentally determined quantities are reported as black solid lines, with shaded areas corresponding to the experimental uncertainty (for $F$ the uncertainty originates from the finite frequency due to the limited number of periods in the the measured Shubnikov-de Haas oscillations; for the masses, we take as a measure of the uncertainty the few percent difference in mass values extracted from the three different devices).}
    \label{fig:Fig4}
\end{figure*}

The information about the extremal Fermi surfaces extracted from the analysis of the SdH oscillations are largely consistent with the properties of the electron pockets of bulk jacutingaite computed by density-functional theory (DFT; simulations were done using the Quantum ESPRESSO distribution\cite{giannozzi_quantum_2009,giannozzi_qe_2017}; see SI for details). When SOI is neglected (see \autoref{fig:Fig1}b), a degeneracy between the top valence and bottom conduction bands is enforced along the vertical edges of the Brillouin zone by the 3-fold rotation symmetry of the crystal (see the KH line in \autoref{fig:Fig1}b), and gives rise to nodal lines. Additional crossings between valence and conduction bands occur away from high-symmetry directions as a consequence of a band inversion driven by the strong interlayer coupling\cite{theory_jacu_2019} (see the $\mathrm{\tilde L}\mathrm{\tilde A}\mathrm{\tilde L}$ line in \autoref{fig:Fig1}b), forming two additional nodal loops\cite{bansil_arxiv_2019,theory_jacu_2019}. The location within the Brillouin zone of these nodal lines and nodal loops is reported in \autoref{fig:Fig4}a. The color coding denotes the common energy of the valence and conduction bands along the nodes relative to the Fermi level (blue = above; red = below), with the dispersion resulting in multiple hole and electron pockets.

To enable comparison with experiments we perform additional DFT simulations including SOI effects.  The nodal lines and loops become slightly gapped, but the overall qualitative picture does not change. \autoref{fig:Fig4}b shows the corresponding Fermi surface, where electron and hole pockets are represented in blue and grey respectively, in good agreement (in terms of number and polarity) with the expectations from \autoref{fig:Fig4}a. We identify four symmetry-inequivalent electron pockets and two pockets with hole character. The extremal contours of these pockets orthogonal to the magnetic field direction ($z$-axis) are also reported, with solid  (dashed) lines corresponding to positive (negative) values of $k_z$. Their projection on the $(k_x,k_y)$ plane is shown in \autoref{fig:Fig4}c, with the same color coding as in panel b. This projection is in very good agreement with recent ARPES measurements\cite{Cucchi2019}, with the only difference that the two types of hole pockets are merged into one in experiments. This qualitative feature is not captured by DFT, unless the Fermi energy is shifted on the hole side by few meVs (see panel d).

The area $S$ of the extremal contours of the Fermi surface that we identity provides an estimate for the expected SdH oscillation frequencies through the Onsager relation\cite{Onsager1952} $F = \hbar/(2\pi e) S$, while the cyclotron mass can be computed through the derivative $m_c = \hbar^2/(2\pi) \partial S/\partial E_F$ with respect to the Fermi energy\cite{Schoenberg}. For electron pockets, in \autoref{fig:Fig4}e we show the predicted values of $F$ and $m_c$ as a function of the shift in Fermi energy with respect to the DFT reference value, obtained for structures relaxed using two different vdW-compliant functionals (the shift in Fermi energy --over a small range, see SI for a quantitative estimate of the corresponding changes in electron/hole densities-- is meant to account either for the presence of small doping due to slight deviations from perfect stoichiometry for an unprecise quantitative prediction of the semimetallic band overlap in DFT ). The calculated frequencies corresponding to the four inequivalent electron pockets are in the same range as the experimentally determined values (gray horizontal lines). Similar agreement is found for $m_c$ (for both functionals the calculated cyclotron mass of one pocket --green circles-- falls outside the measured range by approximately 30 \%). For holes, we find that, when the Fermi energy is shifted so that the hole pockets merge (see \autoref{fig:Fig4}d) as seen in ARPES, the overall cyclotron mass becomes very large, on the order of the bare electron mass, resulting in a strong suppression of magnetic oscillations associated with holes. This explains why in the experiments no frequencies corresponding to hole SdH oscillations are detected. We conclude that both functionals show an overall reasonable agreement with the experiments, and correctly capture the magnitude the observed SdH oscillation frequencies and of the corresponding cyclotron masses; deviations are small and concern details beyond what can be expected from this first experimental work on the transport properties of jacutingaite.

\section{Conclusion}
In summary, the results presented here demonstrate the possibility to realize nano-structured devices based on exfoliated jacutingaite (Pt$_2$HgSe$_3$) crystals that enable the observation of high-quality transport properties. The observed electron mobility values are comparable to those reported in WTe$_2$ layers of the same thickness. This is an extremely promising result, given that our jacutingaite devices are based on crystals grown by using a newly developed process that offers considerable margins of improvement. Crystal growth under high pressure occurs within a short time (hours) and imposes large mechanical constraints to the growing crystals. Fine tuning the growth parameters and making the growth time longer is expected to provide larger crystals of even higher quality, and to allow much larger exfoliated layers to be produced, eventually enabling the realization of monolayer devices. Given that all experiments performed until now --both the transport measurements discussed here and ARPES experiments recently reported-- have shown an overall excellent agreement with the predictions of first-principles calculations. We therefore expect that jacutingaite monolayer devices will represent an extremely promising platform to investigate transport in 2D quantum spin Hall insulators, for which the experiments reported here represent and important and necessary first step.

\begin{acknowledgements}
We sincerely acknowledge Alexandre Ferreira for technical support and Antimo Marrazzo for useful discussions. A.F.M. gratefully acknowledges financial support from the Swiss National Science Foundation (Division II) and from the EU Graphene Flagship project. M.G.\ and M.F. acknowledge support from the Swiss National Science Foundation through the Ambizione program (grants PZ00P2\_174056 and PZ00P2\_174038). Simulation time was provided by CSCS on Piz Daint (project IDs s825 and s917).
\end{acknowledgements}

%

%%%%%%%%%%%%%%%%%%%
%% SUPPLEMENTARY %%
%%%%%%%%%%%%%%%%%%%
\clearpage

\onecolumngrid

\section*{\LARGE\bfseries Supporting Information}

\renewcommand\theequation{S\arabic{equation}}
\renewcommand\thefigure{S\arabic{figure}}
\renewcommand\thetable{S\arabic{figure}}
\renewcommand\thesection{S\arabic{section}}
\renewcommand\thesubsection{S\arabic{section}.\arabic{subsection}}
\setcounter{equation}{0}
\setcounter{figure}{0}
\setcounter{table}{0}
\setcounter{section}{0}

\section{Details on crystal growth}

Because of the high vapour pressure of mercury at low temperatures and the high reaction temperature needed for alloying platinum, the growth of $\mathrm{Pt_2HgSe_3}$ must be carried out under extreme conditions. Single crystals of jacutingaite $\mathrm{Pt_2HgSe_3}$ have been grown from the melt under high isostatic pressure in a hot-stage cubic-anvil press. The precursor components Pt (sponge, 99.98\%), $\mathrm{HgSe}$ (powder, 99.9\%) and Se (shot, 99.999\%) have been mixed with a nominal slightly Hg-exceeding composition $\mathrm{Pt_2Hg_{1.1}Se_3}$ inside a Ar-filled glove box, and pelletized to a relative mass density of $>75\%$. The pellet was tightly inserted into a BN crucible, covered by BN caps, and placed inside a cylinder-like graphite heater. The heater fits inside a pyrophillite cubic cell, the pressure transmitter, on each face of which a WC piston acts, thus transforming a uni-axial force in a quasi-isostatic pressure on the sample. With this apparatus, we could process the material at temperatures of 800-1000 $^{\circ}$C under a pressure of 1-2~GPa. After quite a short heating and cooling process (1-2h at the maximum temperature followed by cooling at 50-70 $^{\circ}$C/h down to 650$^{\circ}$C), crystals of $\mathrm{Pt_2HgSe_3}$ have grown inside the pellet and could be mechanically extracted.\par

The largest size of the crystals was $ \sim 0.6 \times 1.0 \mathrm{mm^2}$ in the cleavage plane (perpendicular to the crystallographic c-axis). Owing to the very high density of the as-grown polycrystalline bulk, the crystal size is limited and the extraction of the crystal is a delicate operation. However, the use of high pressure strongly improves the growth kinetics and allows obtaining bulk crystals within few hours, while the vapor-solid reaction technique reported\cite{jacutingaite_exp_12} could only produce powder samples over a time scale of 70 days.\par

 Single crystals were mechanically extracted from the as-grown boule and individually checked for structural and chemical quality, by means of X-ray diffraction (XRD) and Wavelength Dispersive X-ray Spectroscopy (WDS). The single-crystal XRD confirmed the crystal structure previously reported for jacutingaite $\mathrm{Pt_2HgSe_3}$ \cite{jacutingaite_exp_12}. WDS chemical analysis was performed with a Oxford Aztec Advance complete WDS system coupled to a SDD detector X-MaxN80, installed in a JEOL JSM 7600 F scanning electron microscope. Factory-provided standards Pt, Se and HgTe were used for quantitative calibration. The Pt:Hg ratio was found to be equal to $1.95 \pm 0.05$, averaging over various crystals, thus confirming that no Hg vacancies are present in these crystals at an appreciable amount. Accordingly, the Hg-site in the crystal structure is found to be fully occupied by the Rietveld profile refinement of powder XRD diffraction patterns. 

Part of the as-grown pellet was ground and used for powder X-ray diffraction. Reflections from the secondary phase $\mathrm{PtSe_2}$ were often present in the powder diffraction pattern (up to 6-7\% in volume) and rare single crystals of $\mathrm{PtSe_2}$ were found as well, with a similar size and aspect to those of $\mathrm{Pt_2HgSe_3}$. In fact, $\mathrm{PtSe_2}$ can stably form in the same temperature range as $\mathrm{Pt_2HgSe_3}$ and, because of its structural similarity, can even grow epitaxially next to a $\mathrm{Pt_2HgSe_3}$ grain, as already reported\cite{Drabek2012}. For this reason, it was mandatory to analyze carefully every single crystal prior to proceeding with device preparation and physical measurements.

\begin{figure*}[ht]
  \includegraphics[width= 0.45\textwidth]{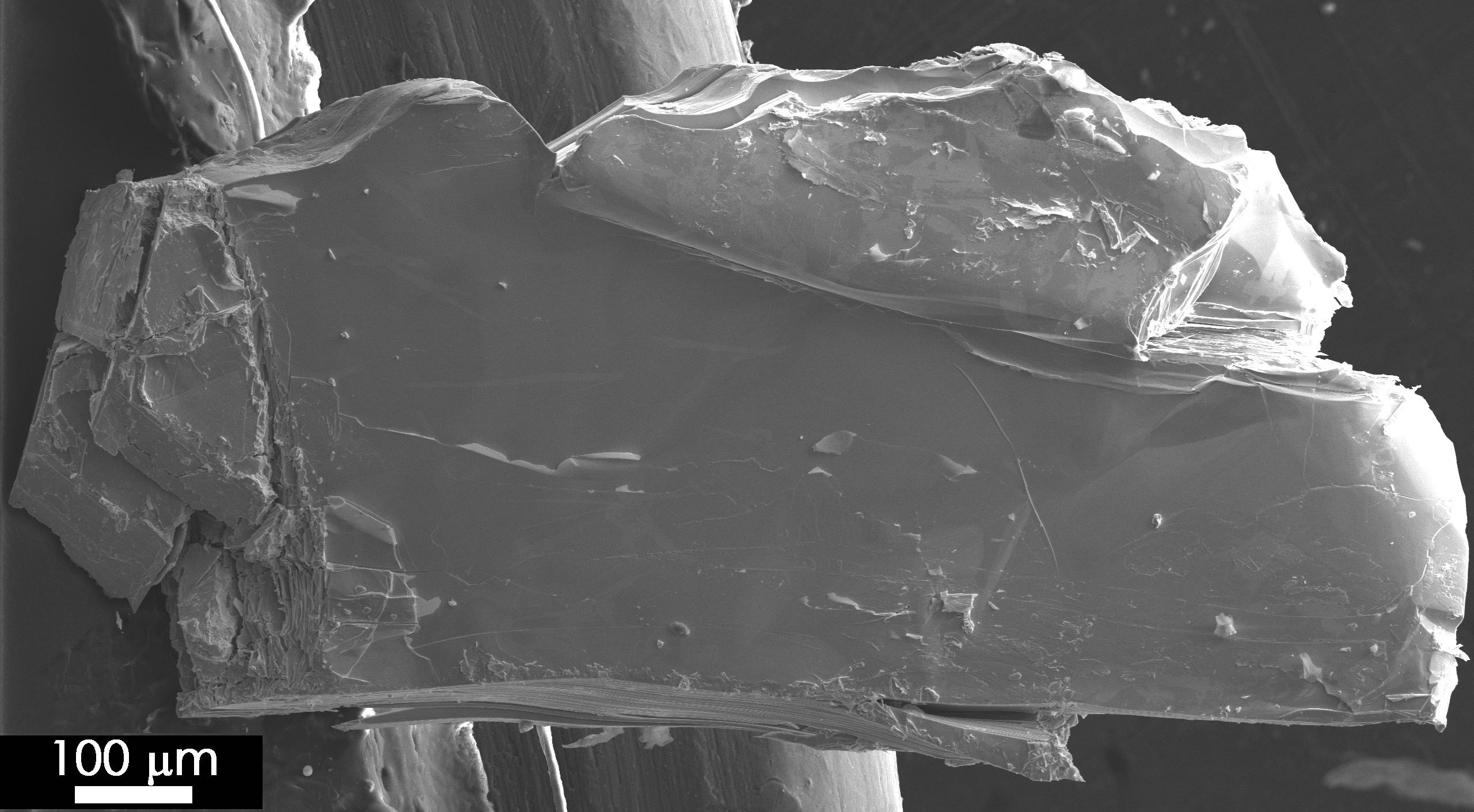}
  \caption{ SEM image of a $\mathrm{Pt_2HgSe_3}$ crystal extracted from the as-grown polycrystalline boule. } 
  \label{fig:SEM}
\end{figure*}

\newpage

\section{Details of first-principles calculations}
First-principles simulations have been carried out within density-functional theory using the Quantum ESPRESSO distribution~\cite{giannozzi_quantum_2009,giannozzi_qe_2017}. Several van-der-Waals compliant functionals have been adopted to relax the atomic positions until the residual force on each atoms was smaller than 2.5~mev/\AA, while keeping the unit cell fixed to the experimental one. At this stage spin-orbit coupling has been neglected and electron-ion interactions have been approximated using pseudopotentials from the Standard Solid State pseudopotential library\cite{prandini_precision_2018} (v1.0), with a cutoff of 50~Ry on energy and 280~Ry on the charge density. Spin-orbit coupling is then included in calculations on the relaxed structure by using fully relativistic pseudopotentials from the Pseudo-Dojo library\cite{dojo_paper_18} with a energy cutoff of 80~Ry. The Brillouin zone is sampled using a $8\times8\times8$ $\Gamma$-centered Monkhorst-Pack grid with a cold smearing\cite{mv_smearing_99} of 0.015~Ry.  To compute the Fermi surface, energy bands are interpolated by mapping the electronic states onto a set of maximally-localized Wannier functions\cite{wannier_review_12} using WANNIER90\cite{mostofi_updated_2014}. As starting projections we consider $s$- and $d$-orbitals on Hg and Pt, together with $p$-orbitals on Se, and the wannierization is performed on a $6\times6\times6$ regular grid over the Brillouin zone. The Fermi surface is computed from energy bands Wannier interpolated over a dense $80\times80\times80$ k-grid. The SKEAF software\cite{Rourke2012} is then adopted to extract the extremal contours, their areas, and the corresponding cyclotron mass. Wannier functions are used also to obtain the reciprocal space dispersion of the nodal loops with the help of the WannierTools software\cite{wannier_tools_18}.

\section{Computed frequencies of Raman active modes}

\begin{figure}[h!]
    \centering
    \includegraphics{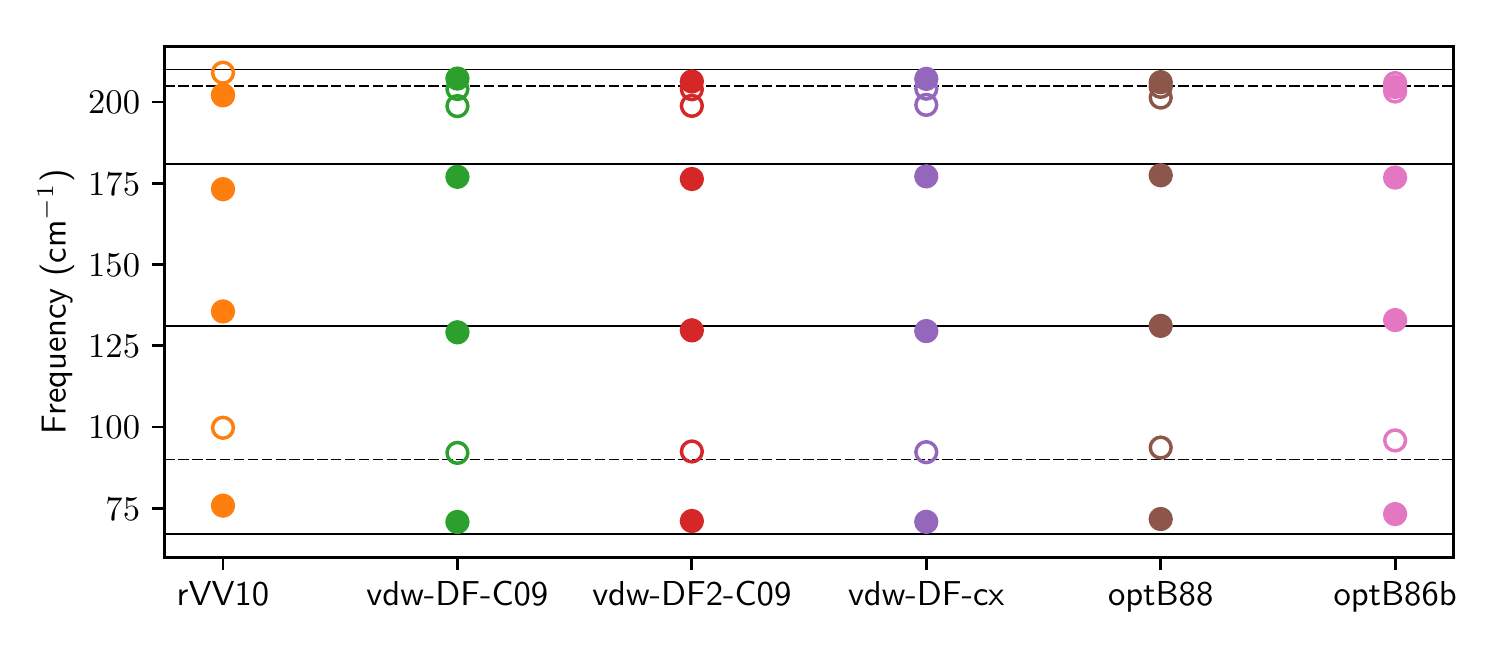}
    \caption{Frequencies (in cm$^{-1}$) of the Raman active modes of bulk jacutingaite computed with different van der Waals functionals. Full and empty symbols correspond to modes belonging to the E$_g$ and A$_{1g}$ representation, respectively. Experimental results are reported as horizontal lines with solid and dashed lines denoting E$_g$ and A$_{1g}$ modes, as identified by polarization-resolved Raman spectroscopy.  An overall good agreement (both in terms of frequencies and symmetry assignment) is obtained with the vdw-DF-cx functional\cite{Lee2010,Berland2014} adopted in the main text.}
    \label{fig:comparison}
\end{figure}

\begin{table}[h!]
\caption{Frequencies  (in cm$^{-1}$) of the phonon modes of bulk jacutingaite at $\Gamma$, together with the symbol of the  irreducible representation they belong to, computed through a finite-difference approach using phonopy~\cite{phonopy}. Different van-der-Waals compliant exchange-correlation functionals have been considered. Bold numbers and symbols correspond to Raman active modes. }
\begin{tabularx}{\linewidth}{  >{\centering\arraybackslash}X  >{\centering\arraybackslash}X  >{\centering\arraybackslash}X  >{\centering\arraybackslash}X  >{\centering\arraybackslash}X  >{\centering\arraybackslash}X  >{\centering\arraybackslash}X  >{\centering\arraybackslash}X  >{\centering\arraybackslash}X  >{\centering\arraybackslash}X  >{\centering\arraybackslash}X  >{\centering\arraybackslash}X }
\toprule
\multicolumn{2}{c}{  rVV10 \cite{Vydrov2010,Sabatini2013} } & \multicolumn{2}{c}{ vdw-DF-c09 \cite{Dion2004,Cooper2010} } & \multicolumn{2}{c}{ vdw-DF2-c09 \cite{Lee2010,Cooper2010} } & \multicolumn{2}{c}{  vdw-DF-cx \cite{Berland2014}} & \multicolumn{2}{c}{  optB86b \cite{Klimes2011} } & \multicolumn{2}{c}{  optB88 \cite{Klimes2009}}\\
\midrule
 65.8 & E$_{u}$ & 62.6 & E$_{u}$ & 62.8 & E$_{u}$ & 62.7 & E$_{u}$ & 63.5 & E$_{u}$ & 64.8 & E$_{u}$ \\
 \textbf{75.8} & \textbf{E$_{g}$} & \textbf{70.8} & \textbf{E$_{g}$} & \textbf{71.1} & \textbf{E$_{g}$} & \textbf{70.8} & \textbf{E$_{g}$} & \textbf{71.7} & \textbf{E$_{g}$} & \textbf{73.2} & \textbf{E$_{g}$} \\
 91.2 & A$_{1u}$ & 86.3 & A$_{2u}$ & 86.6 & A$_{2u}$ & 86.6 & A$_{2u}$ & 88.0 & A$_{2u}$ & 89.9 & A$_{2u}$ \\
 91.4 & A$_{2u}$ & 92.0 & A$_{1u}$ & 91.6 & A$_{1u}$ & 91.9 & A$_{2u}$ & 92.0 & A$_{1u}$ & 92.4 & A$_{1u}$ \\
 93.4 & A$_{2u}$ & \textbf{92.0} & \textbf{A$_{1g}$} & 91.8 & A$_{2u}$ & 91.9 & A$_{1u}$ & 92.3 & A$_{2u}$ & 93.4 & A$_{2u}$ \\
 93.7 & E$_{u}$ & 92.0 & A$_{2u}$ & 91.9 & E$_{u}$ & 92.2 & E$_{u}$ & 93.0 & E$_{u}$ & 94.0 & E$_{u}$ \\
 \textbf{99.8} & \textbf{A$_{1g}$} & 92.1 & E$_{u}$ & \textbf{92.5} & \textbf{A$_{1g}$} & \textbf{92.2} & \textbf{A$_{1g}$} & \textbf{93.7} & \textbf{A$_{1g}$} & \textbf{95.9} & \textbf{A$_{1g}$} \\
 118.7 & E$_{u}$ & 115.8 & E$_{u}$ & 115.8 & E$_{u}$ & 115.8 & E$_{u}$ & 116.3 & E$_{u}$ & 117.3 & E$_{u}$ \\
 \textbf{135.6} & \textbf{E$_{g}$} & \textbf{129.1} & \textbf{E$_{g}$} & \textbf{129.7} & \textbf{E$_{g}$} & \textbf{129.5} & \textbf{E$_{g}$} & \textbf{131.1} & \textbf{E$_{g}$} & \textbf{132.9} & \textbf{E$_{g}$} \\
 140.4 & E$_{u}$ & 140.0 & E$_{u}$ & 139.9 & E$_{u}$ & 140.1 & E$_{u}$ & 140.4 & E$_{u}$ & 140.7 & E$_{u}$ \\
 152.1 & A$_{2u}$ & 146.6 & A$_{2u}$ & 146.7 & A$_{2u}$ & 146.6 & A$_{2u}$ & 147.5 & A$_{2u}$ & 149.3 & A$_{2u}$ \\
 \textbf{173.2} & \textbf{E$_{g}$} & \textbf{177.0} & \textbf{E$_{g}$} & \textbf{176.3} & \textbf{E$_{g}$} & \textbf{177.1} & \textbf{E$_{g}$} & \textbf{177.4} & \textbf{E$_{g}$} & \textbf{176.7} & \textbf{E$_{g}$} \\
 176.5 & A$_{2g}$ & 184.1 & A$_{2g}$ & 183.1 & A$_{2g}$ & 184.0 & A$_{2g}$ & 182.6 & A$_{2g}$ & 180.4 & A$_{2g}$ \\
 201.0 & E$_{u}$ & 196.5 & E$_{u}$ & 196.8 & E$_{u}$ & 196.9 & E$_{u}$ & 198.3 & E$_{u}$ & 200.0 & E$_{u}$ \\
 \textbf{202.1} & \textbf{E$_{g}$} & \textbf{198.7} & \textbf{A$_{1g}$} & \textbf{198.8} & \textbf{A$_{1g}$} & \textbf{199.1} & \textbf{A$_{1g}$} & \textbf{201.4} & \textbf{A$_{1g}$} & \textbf{203.2} & \textbf{A$_{1g}$} \\
 \textbf{202.2} & \textbf{A$_{1g}$} & 202.7 & A$_{2u}$ & 202.9 & A$_{2u}$ & 202.9 & A$_{2u}$ & 203.9 & A$_{2u}$ & \textbf{204.6} & \textbf{E$_{g}$} \\
 207.8 & A$_{2u}$ & \textbf{204.0} & \textbf{A$_{1g}$} & \textbf{203.9} & \textbf{A$_{1g}$} & \textbf{204.1} & \textbf{A$_{1g}$} & \textbf{204.7} & \textbf{A$_{1g}$} & 205.1 & A$_{2u}$ \\
 \textbf{209.0} & \textbf{A$_{1g}$} & \textbf{207.2} & \textbf{E$_{g}$} & \textbf{206.3} & \textbf{E$_{g}$} & \textbf{207.1} & \textbf{E$_{g}$} & \textbf{206.1} & \textbf{E$_{g}$} & \textbf{205.8} & \textbf{A$_{1g}$} \\
 214.1 & A$_{2u}$ & 219.6 & A$_{2u}$ & 218.9 & A$_{2u}$ & 219.6 & A$_{2u}$ & 218.9 & A$_{2u}$ & 217.4 & A$_{2u}$ \\
 218.5 & E$_{u}$ & 221.6 & E$_{u}$ & 221.1 & E$_{u}$ & 221.7 & E$_{u}$ & 221.3 & E$_{u}$ & 220.2 & E$_{u}$ \\
 220.2 & E$_{u}$ & 225.3 & E$_{u}$ & 224.5 & E$_{u}$ & 225.2 & E$_{u}$ & 224.0 & E$_{u}$ & 222.1 & E$_{u}$ \\
 225.9 & A$_{1u}$ & 230.3 & A$_{1u}$ & 229.6 & A$_{1u}$ & 230.3 & A$_{1u}$ & 229.3 & A$_{1u}$ & 227.8 & A$_{1u}$ \\
\bottomrule
\end{tabularx}
\end{table}

\newpage
\section{Carrier density evolution with Fermi energy shift}

\begin{figure}[h!]
    \centering
    \includegraphics{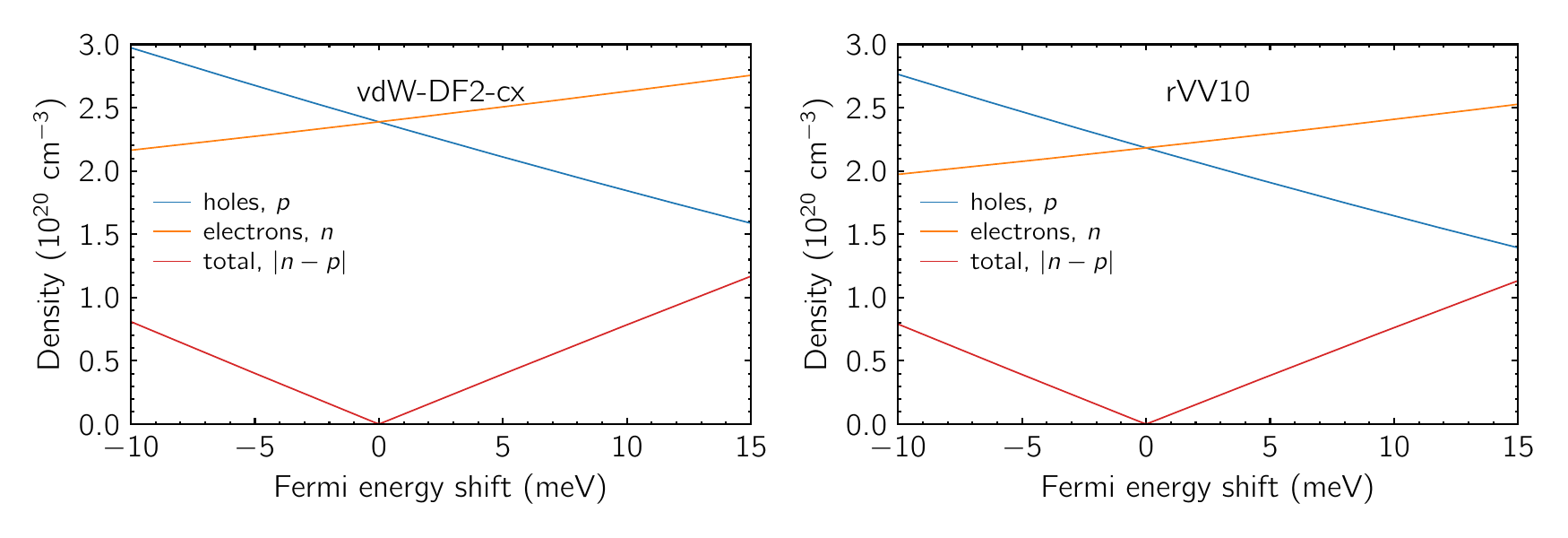}
    \caption{Calculated evolution of the density of electrons ($n$, orange solid line) and holes ($p$, blue solid line) as a function of the Fermi energy shift with respect to the reference value corresponding to overall charge neutrality ($n=p$). The absolute value of the total carrier density is also reported (red solid line). The left and right panels refer to structures relaxed using two different van-der-Waals compliant exchange-correlation functionals within density-functional theory: vdw-DF-cx~\cite{Berland2014} (left) and rVV10~\cite{Vydrov2010,Sabatini2013} (right).}
\end{figure}


\begin{thebibliography}{10}%
\makeatletter
\providecommand \@ifxundefined [1]{%
 \ifx #1\undefined \expandafter \@firstoftwo
 \else \expandafter \@secondoftwo
\fi
}%
\providecommand \@ifnum [1]{%
 \ifnum #1\expandafter \@firstoftwo
 \else \expandafter \@secondoftwo
\fi
}%
\providecommand \enquote [1]{``#1''}%
\providecommand \bibnamefont  [1]{#1}%
\providecommand \bibfnamefont [1]{#1}%
\providecommand \citenamefont [1]{#1}%
\providecommand\href[0]{\@sanitize\@href}%
\providecommand\@href[1]{\endgroup\@@startlink{#1}\endgroup\@@href}%
\providecommand\@@href[1]{#1\@@endlink}%
\providecommand \@sanitize [0]{\begingroup\catcode`\&12\catcode`\#12\relax}%
\@ifxundefined \pdfoutput {\@firstoftwo}{%
 \@ifnum{\z@=\pdfoutput}{\@firstoftwo}{\@secondoftwo}%
}{%
 \providecommand\@@startlink[1]{\leavevmode}%
 \providecommand\@@endlink[0]{}%
}{%
 \providecommand\@@startlink[1]{%
  \leavevmode
  \pdfstartlink
   attr{/Border[0 0 1 ]/H/I/C[0 1 1]}%
   user{/Subtype/Link/A<</Type/Action/S/URI/URI(#1)>>}%
  \relax
 }%
 \providecommand\@@endlink[0]{\pdfendlink}%
}%
\providecommand \url  [0]{\begingroup\@sanitize \@url }%
\providecommand \@url [1]{\endgroup\@href {#1}{\urlprefix}}%
\providecommand \urlprefix [0]{URL }%
\providecommand \Eprint[0]{\href }%
\@ifxundefined \urlstyle {%
  \providecommand \doi [1]{doi:\discretionary{}{}{}#1}%
}{%
  \providecommand \doi [0]{doi:\discretionary{}{}{}\begingroup
  \urlstyle{rm}\Url }%
}%
\providecommand \doibase [0]{http://dx.doi.org/}%
\providecommand \Doi[1]{\href{\doibase#1}}%
\providecommand \bibAnnote [3]{%
  \BibitemShut{#1}%
  \begin{quotation}\noindent
    \textsc{Key:}\ #2\\\textsc{Annotation:}\ #3%
  \end{quotation}%
}%
\providecommand \bibAnnoteFile [2]{%
  \IfFileExists{#2}{\bibAnnote {#1} {#2} {\input{#2}}}{}%
}%
\providecommand \typeout [0]{\immediate \write \m@ne }%
\providecommand \selectlanguage [0]{\@gobble}%
\providecommand \bibinfo [0]{\@secondoftwo}%
\providecommand \bibfield [0]{\@secondoftwo}%
\providecommand \translation [1]{[#1]}%
\providecommand \BibitemOpen[0]{}%
\providecommand \bibitemStop [0]{}%
\providecommand \bibitemNoStop [0]{.\EOS\space}%
\providecommand \EOS [0]{\spacefactor3000\relax}%
\providecommand \BibitemShut [1]{\csname bibitem#1\endcsname}%
%</preamble>
\bibitem{kane_quantum_2005}%
  \BibitemOpen
  \bibfield{author}{%
  \bibinfo {author} {\bibfnamefont{C.~L.}\ \bibnamefont{Kane}}\ and\ \bibinfo
  {author} {\bibfnamefont{E.~J.}\ \bibnamefont{Mele}},\ }%
  \Doi{10.1103/PhysRevLett.95.226801}{\emph{\bibinfo {title} {Quantum {{Spin
  Hall Effect}} in {{Graphene}}}}},\ \bibinfo {journal} {Physical Review
  Letters}\ \textbf{\bibinfo {volume} {95}},\ \bibinfo {pages} {226801}
  (\bibinfo {year} {2005}).~%
  \bibAnnoteFile{Stop}{kane_quantum_2005}%
\bibitem{kane_z2_05}%
  \BibitemOpen
  \bibfield{author}{%
  \bibinfo {author} {\bibfnamefont{C.~L.}\ \bibnamefont{Kane}}\ and\ \bibinfo
  {author} {\bibfnamefont{E.~J.}\ \bibnamefont{Mele}},\ }%
  \Doi{10.1103/PhysRevLett.95.146802}{\emph{\bibinfo {title} {$\mathbb{Z}_2$
  Topological Order and the Quantum Spin Hall Effect}}},\ \bibinfo {journal}
  {Physical Review Letters}\ \textbf{\bibinfo {volume} {95}},\ \bibinfo {pages}
  {146802} (\bibinfo {year} {2005}).~%
  \bibAnnoteFile{Stop}{kane_z2_05}%
\bibitem{Bernevig2006}%
  \BibitemOpen
  \bibfield{author}{%
  \bibinfo {author} {\bibfnamefont{B.~A.}\ \bibnamefont{Bernevig}}\ and\
  \bibinfo {author} {\bibfnamefont{S.-C.}\ \bibnamefont{Zhang}},\ }%
  \Doi{10.1103/PhysRevLett.96.106802}{\emph{\bibinfo {title} {Quantum Spin Hall
  Effect}}},\ \bibinfo {journal} {Phys. Rev. Lett.}\ \textbf{\bibinfo {volume}
  {96}},\ \bibinfo {pages} {106802} (\bibinfo {year} {2006}).~%
  \bibAnnoteFile{Stop}{Bernevig2006}%
\bibitem{Konig2007}%
  \BibitemOpen
  \bibfield{author}{%
  \bibinfo {author} {\bibfnamefont{M.}~\bibnamefont{K{\"o}nig}}, \bibinfo
  {author} {\bibfnamefont{S.}~\bibnamefont{Wiedmann}}, \bibinfo {author}
  {\bibfnamefont{C.}~\bibnamefont{Br{\"u}ne}}, \bibinfo {author}
  {\bibfnamefont{A.}~\bibnamefont{Roth}}, \bibinfo {author}
  {\bibfnamefont{H.}~\bibnamefont{Buhmann}}, \bibinfo {author}
  {\bibfnamefont{L.~W.}\ \bibnamefont{Molenkamp}}, \bibinfo {author}
  {\bibfnamefont{X.-L.}\ \bibnamefont{Qi}},\ and\ \bibinfo {author}
  {\bibfnamefont{S.-C.}\ \bibnamefont{Zhang}},\ }%
  \Doi{10.1126/science.1148047}{\emph{\bibinfo {title} {Quantum Spin Hall
  Insulator State in HgTe Quantum Wells}}},\ \bibinfo {journal} {Science}\
  \textbf{\bibinfo {volume} {318}},\ \bibinfo {pages} {766} (\bibinfo {year}
  {2007}).~%
  \bibAnnoteFile{Stop}{Konig2007}%
\bibitem{Knez2011}%
  \BibitemOpen
  \bibfield{author}{%
  \bibinfo {author} {\bibfnamefont{I.}~\bibnamefont{Knez}}, \bibinfo {author}
  {\bibfnamefont{R.~R.}\ \bibnamefont{Du}},\ and\ \bibinfo {author}
  {\bibfnamefont{G.}~\bibnamefont{Sullivan}},\ }%
  \Doi{10.1103/PhysRevLett.107.136603}{\emph{\bibinfo {title} {{Evidence for
  helical edge modes in inverted InAs/GaSb quantum wells}}}},\ \bibinfo
  {journal} {Physical Review Letters}\ \textbf{\bibinfo {volume} {107}},\
  \bibinfo {pages} {1} (\bibinfo {year} {2011}).~%
  \bibAnnoteFile{Stop}{Knez2011}%
\bibitem{Suzuki2013}%
  \BibitemOpen
  \bibfield{author}{%
  \bibinfo {author} {\bibfnamefont{K.}~\bibnamefont{Suzuki}}, \bibinfo {author}
  {\bibfnamefont{Y.}~\bibnamefont{Harada}}, \bibinfo {author}
  {\bibfnamefont{K.}~\bibnamefont{Onomitsu}},\ and\ \bibinfo {author}
  {\bibfnamefont{K.}~\bibnamefont{Muraki}},\ }%
  \Doi{10.1103/PhysRevB.87.235311}{\emph{\bibinfo {title} {{Edge channel
  transport in the InAs/GaSb topological insulating phase}}}},\ \bibinfo
  {journal} {Physical Review B - Condensed Matter and Materials Physics}\
  \textbf{\bibinfo {volume} {87}},\ \bibinfo {pages} {1} (\bibinfo {year}
  {2013}).~%
  \bibAnnoteFile{Stop}{Suzuki2013}%
\bibitem{Fei2017}%
  \BibitemOpen
  \bibfield{author}{%
  \bibinfo {author} {\bibfnamefont{Z.}~\bibnamefont{Fei}}, \bibinfo {author}
  {\bibfnamefont{T.}~\bibnamefont{Palomaki}}, \bibinfo {author}
  {\bibfnamefont{S.}~\bibnamefont{Wu}}, \bibinfo {author}
  {\bibfnamefont{W.}~\bibnamefont{Zhao}}, \bibinfo {author}
  {\bibfnamefont{X.}~\bibnamefont{Cai}}, \bibinfo {author}
  {\bibfnamefont{B.}~\bibnamefont{Sun}}, \bibinfo {author}
  {\bibfnamefont{P.}~\bibnamefont{Nguyen}}, \bibinfo {author}
  {\bibfnamefont{J.}~\bibnamefont{Finney}}, \bibinfo {author}
  {\bibfnamefont{X.}~\bibnamefont{Xu}},\ and\ \bibinfo {author}
  {\bibfnamefont{D.~H.}\ \bibnamefont{Cobden}},\ }%
  \Doi{10.1038/nphys4091}{\emph{\bibinfo {title} {{Edge conduction in monolayer
  $\mathrm{WTe}_2$}}}},\ \bibinfo {journal} {Nature Physics}\ \textbf{\bibinfo
  {volume} {13}},\ \bibinfo {pages} {677} (\bibinfo {year} {2017}).~%
  \bibAnnoteFile{Stop}{Fei2017}%
\bibitem{Wu2018}%
  \BibitemOpen
  \bibfield{author}{%
  \bibinfo {author} {\bibfnamefont{S.}~\bibnamefont{Wu}}, \bibinfo {author}
  {\bibfnamefont{V.}~\bibnamefont{Fatemi}}, \bibinfo {author}
  {\bibfnamefont{Q.~D.}\ \bibnamefont{Gibson}}, \bibinfo {author}
  {\bibfnamefont{K.}~\bibnamefont{Watanabe}}, \bibinfo {author}
  {\bibfnamefont{T.}~\bibnamefont{Taniguchi}}, \bibinfo {author}
  {\bibfnamefont{R.~J.}\ \bibnamefont{Cava}},\ and\ \bibinfo {author}
  {\bibfnamefont{P.}~\bibnamefont{Jarillo-Herrero}},\ }%
  \Doi{10.1126/science.aan6003}{\emph{\bibinfo {title} {{Observation of the
  quantum spin Hall effect up to 100 kelvin in a monolayer crystal}}}},\
  \bibinfo {journal} {Science}\ \textbf{\bibinfo {volume} {359}},\ \bibinfo
  {pages} {76} (\bibinfo {year} {2018}).~%
  \bibAnnoteFile{Stop}{Wu2018}%
\bibitem{Roth2009}%
  \BibitemOpen
  \bibfield{author}{%
  \bibinfo {author} {\bibfnamefont{A.}~\bibnamefont{Roth}}, \bibinfo {author}
  {\bibfnamefont{C.}~\bibnamefont{Br{\"u}ne}}, \bibinfo {author}
  {\bibfnamefont{H.}~\bibnamefont{Buhmann}}, \bibinfo {author}
  {\bibfnamefont{L.~W.}\ \bibnamefont{Molenkamp}}, \bibinfo {author}
  {\bibfnamefont{J.}~\bibnamefont{Maciejko}}, \bibinfo {author}
  {\bibfnamefont{X.-L.}\ \bibnamefont{Qi}},\ and\ \bibinfo {author}
  {\bibfnamefont{S.-C.}\ \bibnamefont{Zhang}},\ }%
  \Doi{10.1126/science.1174736}{\emph{\bibinfo {title} {Nonlocal Transport in
  the Quantum Spin Hall State}}},\ \bibinfo {journal} {Science}\
  \textbf{\bibinfo {volume} {325}},\ \bibinfo {pages} {294} (\bibinfo {year}
  {2009}).~%
  \bibAnnoteFile{Stop}{Roth2009}%
\bibitem{Grabecki2013}%
  \BibitemOpen
  \bibfield{author}{%
  \bibinfo {author} {\bibfnamefont{G.}~\bibnamefont{Grabecki}}, \bibinfo
  {author} {\bibfnamefont{J.}~\bibnamefont{Wr\'obel}}, \bibinfo {author}
  {\bibfnamefont{M.}~\bibnamefont{Czapkiewicz}}, \bibinfo {author}
  {\bibfnamefont{L.}~\bibnamefont{Cywi\ifmmode~\acute{n}\else \'{n}\fi{}ski}},
  \bibinfo {author} {\bibfnamefont{S.}~\bibnamefont{Giera\l{}towska}}, \bibinfo
  {author} {\bibfnamefont{E.}~\bibnamefont{Guziewicz}}, \bibinfo {author}
  {\bibfnamefont{M.}~\bibnamefont{Zholudev}}, \bibinfo {author}
  {\bibfnamefont{V.}~\bibnamefont{Gavrilenko}}, \bibinfo {author}
  {\bibfnamefont{N.~N.}\ \bibnamefont{Mikhailov}}, \bibinfo {author}
  {\bibfnamefont{S.~A.}\ \bibnamefont{Dvoretski}}, \bibinfo {author}
  {\bibfnamefont{F.}~\bibnamefont{Teppe}}, \bibinfo {author}
  {\bibfnamefont{W.}~\bibnamefont{Knap}},\ and\ \bibinfo {author}
  {\bibfnamefont{T.}~\bibnamefont{Dietl}},\ }%
  \Doi{10.1103/PhysRevB.88.165309}{\emph{\bibinfo {title} {Nonlocal resistance
  and its fluctuations in microstructures of band-inverted HgTe/(Hg,Cd)Te
  quantum wells}}},\ \bibinfo {journal} {Phys. Rev. B}\ \textbf{\bibinfo
  {volume} {88}},\ \bibinfo {pages} {165309} (\bibinfo {year} {2013}).~%
  \bibAnnoteFile{Stop}{Grabecki2013}%
\bibitem{Konig2013}%
  \BibitemOpen
  \bibfield{author}{%
  \bibinfo {author} {\bibfnamefont{M.}~\bibnamefont{K\"onig}}, \bibinfo
  {author} {\bibfnamefont{M.}~\bibnamefont{Baenninger}}, \bibinfo {author}
  {\bibfnamefont{A.~G.~F.}\ \bibnamefont{Garcia}}, \bibinfo {author}
  {\bibfnamefont{N.}~\bibnamefont{Harjee}}, \bibinfo {author}
  {\bibfnamefont{B.~L.}\ \bibnamefont{Pruitt}}, \bibinfo {author}
  {\bibfnamefont{C.}~\bibnamefont{Ames}}, \bibinfo {author}
  {\bibfnamefont{P.}~\bibnamefont{Leubner}}, \bibinfo {author}
  {\bibfnamefont{C.}~\bibnamefont{Br\"une}}, \bibinfo {author}
  {\bibfnamefont{H.}~\bibnamefont{Buhmann}}, \bibinfo {author}
  {\bibfnamefont{L.~W.}\ \bibnamefont{Molenkamp}},\ and\ \bibinfo {author}
  {\bibfnamefont{D.}~\bibnamefont{Goldhaber-Gordon}},\ }%
  \Doi{10.1103/PhysRevX.3.021003}{\emph{\bibinfo {title} {Spatially Resolved
  Study of Backscattering in the Quantum Spin Hall State}}},\ \bibinfo
  {journal} {Phys. Rev. X}\ \textbf{\bibinfo {volume} {3}},\ \bibinfo {pages}
  {021003} (\bibinfo {year} {2013}).~%
  \bibAnnoteFile{Stop}{Konig2013}%
\bibitem{Wu2006}%
  \BibitemOpen
  \bibfield{author}{%
  \bibinfo {author} {\bibfnamefont{C.}~\bibnamefont{Wu}}, \bibinfo {author}
  {\bibfnamefont{B.~A.}\ \bibnamefont{Bernevig}},\ and\ \bibinfo {author}
  {\bibfnamefont{S.-C.}\ \bibnamefont{Zhang}},\ }%
  \Doi{10.1103/PhysRevLett.96.106401}{\emph{\bibinfo {title} {Helical Liquid
  and the Edge of Quantum Spin Hall Systems}}},\ \bibinfo {journal} {Phys. Rev.
  Lett.}\ \textbf{\bibinfo {volume} {96}},\ \bibinfo {pages} {106401} (\bibinfo
  {year} {2006}).~%
  \bibAnnoteFile{Stop}{Wu2006}%
\bibitem{Schmidt2011}%
  \BibitemOpen
  \bibfield{author}{%
  \bibinfo {author} {\bibfnamefont{T.~L.}\ \bibnamefont{Schmidt}},\ }%
  \Doi{10.1103/PhysRevLett.107.096602}{\emph{\bibinfo {title} {Current
  Correlations in Quantum Spin Hall Insulators}}},\ \bibinfo {journal} {Phys.
  Rev. Lett.}\ \textbf{\bibinfo {volume} {107}},\ \bibinfo {pages} {096602}
  (\bibinfo {year} {2011}).~%
  \bibAnnoteFile{Stop}{Schmidt2011}%
\bibitem{Budich2012}%
  \BibitemOpen
  \bibfield{author}{%
  \bibinfo {author} {\bibfnamefont{J.~C.}\ \bibnamefont{Budich}}, \bibinfo
  {author} {\bibfnamefont{F.}~\bibnamefont{Dolcini}}, \bibinfo {author}
  {\bibfnamefont{P.}~\bibnamefont{Recher}},\ and\ \bibinfo {author}
  {\bibfnamefont{B.}~\bibnamefont{Trauzettel}},\ }%
  \Doi{10.1103/PhysRevLett.108.086602}{\emph{\bibinfo {title} {Phonon-Induced
  Backscattering in Helical Edge States}}},\ \bibinfo {journal} {Phys. Rev.
  Lett.}\ \textbf{\bibinfo {volume} {108}},\ \bibinfo {pages} {086602}
  (\bibinfo {year} {2012}).~%
  \bibAnnoteFile{Stop}{Budich2012}%
\bibitem{Vayrynen2013}%
  \BibitemOpen
  \bibfield{author}{%
  \bibinfo {author} {\bibfnamefont{J.~I.}\ \bibnamefont{V\"ayrynen}}, \bibinfo
  {author} {\bibfnamefont{M.}~\bibnamefont{Goldstein}},\ and\ \bibinfo {author}
  {\bibfnamefont{L.~I.}\ \bibnamefont{Glazman}},\ }%
  \Doi{10.1103/PhysRevLett.110.216402}{\emph{\bibinfo {title} {Helical Edge
  Resistance Introduced by Charge Puddles}}},\ \bibinfo {journal} {Phys. Rev.
  Lett.}\ \textbf{\bibinfo {volume} {110}},\ \bibinfo {pages} {216402}
  (\bibinfo {year} {2013}).~%
  \bibAnnoteFile{Stop}{Vayrynen2013}%
\bibitem{Wang2017}%
  \BibitemOpen
  \bibfield{author}{%
  \bibinfo {author} {\bibfnamefont{J.}~\bibnamefont{Wang}}, \bibinfo {author}
  {\bibfnamefont{Y.}~\bibnamefont{Meir}},\ and\ \bibinfo {author}
  {\bibfnamefont{Y.}~\bibnamefont{Gefen}},\ }%
  \Doi{10.1103/PhysRevLett.118.046801}{\emph{\bibinfo {title} {Spontaneous
  Breakdown of Topological Protection in Two Dimensions}}},\ \bibinfo {journal}
  {Phys. Rev. Lett.}\ \textbf{\bibinfo {volume} {118}},\ \bibinfo {pages}
  {046801} (\bibinfo {year} {2017}).~%
  \bibAnnoteFile{Stop}{Wang2017}%
\bibitem{Novelli2019}%
  \BibitemOpen
  \bibfield{author}{%
  \bibinfo {author} {\bibfnamefont{P.}~\bibnamefont{Novelli}}, \bibinfo
  {author} {\bibfnamefont{F.}~\bibnamefont{Taddei}}, \bibinfo {author}
  {\bibfnamefont{A.~K.}\ \bibnamefont{Geim}},\ and\ \bibinfo {author}
  {\bibfnamefont{M.}~\bibnamefont{Polini}},\ }%
  \Doi{10.1103/PhysRevLett.122.016601}{\emph{\bibinfo {title} {Failure of
  Conductance Quantization in Two-Dimensional Topological Insulators due to
  Nonmagnetic Impurities}}},\ \bibinfo {journal} {Phys. Rev. Lett.}\
  \textbf{\bibinfo {volume} {122}},\ \bibinfo {pages} {016601} (\bibinfo {year}
  {2019}).~%
  \bibAnnoteFile{Stop}{Novelli2019}%
\bibitem{Qian2014}%
  \BibitemOpen
  \bibfield{author}{%
  \bibinfo {author} {\bibfnamefont{X.}~\bibnamefont{Qian}}, \bibinfo {author}
  {\bibfnamefont{J.}~\bibnamefont{Liu}}, \bibinfo {author}
  {\bibfnamefont{L.}~\bibnamefont{Fu}},\ and\ \bibinfo {author}
  {\bibfnamefont{J.}~\bibnamefont{Li}},\ }%
  \Doi{10.1126/science.1256815}{\emph{\bibinfo {title} {{Quantum spin Hall
  effect in two- dimensional transition metal dichalcogenides}}}},\ \bibinfo
  {journal} {Science express}\ \textbf{\bibinfo {volume} {10}},\ \bibinfo
  {pages} {1126} (\bibinfo {year} {2014}).~%
  \bibAnnoteFile{Stop}{Qian2014}%
\bibitem{Tang2017}%
  \BibitemOpen
  \bibfield{author}{%
  \bibinfo {author} {\bibfnamefont{S.}~\bibnamefont{Tang}}, \bibinfo {author}
  {\bibfnamefont{C.}~\bibnamefont{Zhang}}, \bibinfo {author}
  {\bibfnamefont{D.}~\bibnamefont{Wong}}, \bibinfo {author}
  {\bibfnamefont{Z.}~\bibnamefont{Pedramrazi}}, \bibinfo {author}
  {\bibfnamefont{H.~Z.}\ \bibnamefont{Tsai}}, \bibinfo {author}
  {\bibfnamefont{C.}~\bibnamefont{Jia}}, \bibinfo {author}
  {\bibfnamefont{B.}~\bibnamefont{Moritz}}, \bibinfo {author}
  {\bibfnamefont{M.}~\bibnamefont{Claassen}}, \bibinfo {author}
  {\bibfnamefont{H.}~\bibnamefont{Ryu}}, \bibinfo {author}
  {\bibfnamefont{S.}~\bibnamefont{Kahn}}, \bibinfo {author}
  {\bibfnamefont{J.}~\bibnamefont{Jiang}}, \bibinfo {author}
  {\bibfnamefont{H.}~\bibnamefont{Yan}}, \bibinfo {author}
  {\bibfnamefont{M.}~\bibnamefont{Hashimoto}}, \bibinfo {author}
  {\bibfnamefont{D.}~\bibnamefont{Lu}}, \bibinfo {author}
  {\bibfnamefont{R.~G.}\ \bibnamefont{Moore}}, \bibinfo {author}
  {\bibfnamefont{C.~C.}\ \bibnamefont{Hwang}}, \bibinfo {author}
  {\bibfnamefont{C.}~\bibnamefont{Hwang}}, \bibinfo {author}
  {\bibfnamefont{Z.}~\bibnamefont{Hussain}}, \bibinfo {author}
  {\bibfnamefont{Y.}~\bibnamefont{Chen}}, \bibinfo {author}
  {\bibfnamefont{M.~M.}\ \bibnamefont{Ugeda}}, \bibinfo {author}
  {\bibfnamefont{Z.}~\bibnamefont{Liu}}, \bibinfo {author}
  {\bibfnamefont{X.}~\bibnamefont{Xie}}, \bibinfo {author}
  {\bibfnamefont{T.~P.}\ \bibnamefont{Devereaux}}, \bibinfo {author}
  {\bibfnamefont{M.~F.}\ \bibnamefont{Crommie}}, \bibinfo {author}
  {\bibfnamefont{S.~K.}\ \bibnamefont{Mo}},\ and\ \bibinfo {author}
  {\bibfnamefont{Z.~X.}\ \bibnamefont{Shen}},\ }%
  \Doi{10.1038/nphys4174}{\emph{\bibinfo {title} {{Quantum spin Hall state in
  monolayer $\mathrm{1T'-WTe}_2$}}}},\ \bibinfo {journal} {Nature Physics}\
  \textbf{\bibinfo {volume} {13}},\ \bibinfo {pages} {683} (\bibinfo {year}
  {2017}).~%
  \bibAnnoteFile{Stop}{Tang2017}%
\bibitem{Shi2019}%
  \BibitemOpen
  \bibfield{author}{%
  \bibinfo {author} {\bibfnamefont{Y.}~\bibnamefont{Shi}}, \bibinfo {author}
  {\bibfnamefont{J.}~\bibnamefont{Kahn}}, \bibinfo {author}
  {\bibfnamefont{B.}~\bibnamefont{Niu}}, \bibinfo {author}
  {\bibfnamefont{Z.}~\bibnamefont{Fei}}, \bibinfo {author}
  {\bibfnamefont{B.}~\bibnamefont{Sun}}, \bibinfo {author}
  {\bibfnamefont{X.}~\bibnamefont{Cai}}, \bibinfo {author}
  {\bibfnamefont{B.~A.}\ \bibnamefont{Francisco}}, \bibinfo {author}
  {\bibfnamefont{D.}~\bibnamefont{Wu}}, \bibinfo {author}
  {\bibfnamefont{Z.~X.}\ \bibnamefont{Shen}}, \bibinfo {author}
  {\bibfnamefont{X.}~\bibnamefont{Xu}}, \bibinfo {author}
  {\bibfnamefont{D.~H.}\ \bibnamefont{Cobden}},\ and\ \bibinfo {author}
  {\bibfnamefont{Y.~T.}\ \bibnamefont{Cui}},\ }%
  \Doi{10.1126/sciadv.aat8799}{\emph{\bibinfo {title} {{Imaging quantum spin
  Hall edges in monolayer $\mathrm{WTe}_2$}}}},\ \bibinfo {journal} {Science
  Advances}\ \textbf{\bibinfo {volume} {5}},\ \bibinfo {pages} {1} (\bibinfo
  {year} {2019}).~%
  \bibAnnoteFile{Stop}{Shi2019}%
\bibitem{Cucchi2019}%
  \BibitemOpen
  \bibfield{author}{%
  \bibinfo {author} {\bibfnamefont{I.}~\bibnamefont{Cucchi}}, \bibinfo {author}
  {\bibfnamefont{I.}~\bibnamefont{Guti{\'{e}}rrez-Lezama}}, \bibinfo {author}
  {\bibfnamefont{E.}~\bibnamefont{Cappelli}}, \bibinfo {author}
  {\bibfnamefont{S.~M.~K.}\ \bibnamefont{Walker}}, \bibinfo {author}
  {\bibfnamefont{F.~Y.}\ \bibnamefont{Bruno}}, \bibinfo {author}
  {\bibfnamefont{G.}~\bibnamefont{Tenasini}}, \bibinfo {author}
  {\bibfnamefont{L.}~\bibnamefont{Wang}}, \bibinfo {author}
  {\bibfnamefont{N.}~\bibnamefont{Ubrig}}, \bibinfo {author}
  {\bibfnamefont{C.}~\bibnamefont{Barreteau}}, \bibinfo {author}
  {\bibfnamefont{E.}~\bibnamefont{Giannini}}, \bibinfo {author}
  {\bibfnamefont{M.}~\bibnamefont{Gibertini}}, \bibinfo {author}
  {\bibfnamefont{A.}~\bibnamefont{Tamai}}, \bibinfo {author}
  {\bibfnamefont{A.~F.}\ \bibnamefont{Morpurgo}},\ and\ \bibinfo {author}
  {\bibfnamefont{F.}~\bibnamefont{Baumberger}},\ }%
  \Doi{10.1021/acs.nanolett.8b04534}{\emph{\bibinfo {title} {{Microfocus
  Laser-Angle-Resolved Photoemission on Encapsulated Mono-, Bi-, and Few-Layer
  $\mathrm{1T′-WTe}_2$}}}},\ \bibinfo {journal} {Nano Letters}\
  \textbf{\bibinfo {volume} {19}},\ \bibinfo {pages} {554} (\bibinfo {year}
  {2019}).~%
  \bibAnnoteFile{Stop}{Cucchi2019}%
\bibitem{Zheng2016}%
  \BibitemOpen
  \bibfield{author}{%
  \bibinfo {author} {\bibfnamefont{F.}~\bibnamefont{Zheng}}, \bibinfo {author}
  {\bibfnamefont{C.}~\bibnamefont{Cai}}, \bibinfo {author}
  {\bibfnamefont{S.}~\bibnamefont{Ge}}, \bibinfo {author}
  {\bibfnamefont{X.}~\bibnamefont{Zhang}}, \bibinfo {author}
  {\bibfnamefont{X.}~\bibnamefont{Liu}}, \bibinfo {author}
  {\bibfnamefont{H.}~\bibnamefont{Lu}}, \bibinfo {author}
  {\bibfnamefont{Y.}~\bibnamefont{Zhang}}, \bibinfo {author}
  {\bibfnamefont{J.}~\bibnamefont{Qiu}}, \bibinfo {author}
  {\bibfnamefont{T.}~\bibnamefont{Taniguchi}}, \bibinfo {author}
  {\bibfnamefont{K.}~\bibnamefont{Watanabe}}, \bibinfo {author}
  {\bibfnamefont{S.}~\bibnamefont{Jia}}, \bibinfo {author}
  {\bibfnamefont{J.}~\bibnamefont{Qi}}, \bibinfo {author}
  {\bibfnamefont{J.-H.}\ \bibnamefont{Chen}}, \bibinfo {author}
  {\bibfnamefont{D.}~\bibnamefont{Sun}},\ and\ \bibinfo {author}
  {\bibfnamefont{J.}~\bibnamefont{Feng}},\ }%
  \Doi{10.1002/adma.201600100}{\emph{\bibinfo {title} {On the Quantum Spin Hall
  Gap of Monolayer $\mathrm{1T′-WTe}_2$}}},\ \bibinfo {journal} {Advanced
  Materials}\ \textbf{\bibinfo {volume} {28}},\ \bibinfo {pages} {4845}
  (\bibinfo {year} {2016}).~%
  \bibAnnoteFile{Stop}{Zheng2016}%
\bibitem{dresden_htz2_2018}%
  \BibitemOpen
  \bibfield{author}{%
  \bibinfo {author} {\bibfnamefont{X.}~\bibnamefont{Li}}, \bibinfo {author}
  {\bibfnamefont{Z.}~\bibnamefont{Zhang}}, \bibinfo {author}
  {\bibfnamefont{Y.}~\bibnamefont{Yao}},\ and\ \bibinfo {author}
  {\bibfnamefont{H.}~\bibnamefont{Zhang}},\ }%
  \Doi{10.1088/2053-1583/aadb1e}{\emph{\bibinfo {title} {High Throughput
  Screening for Two-Dimensional Topological Insulators}}},\ \bibinfo {journal}
  {2D Materials}\ \textbf{\bibinfo {volume} {5}},\ \bibinfo {pages} {045023}
  (\bibinfo {year} {2018}).~%
  \bibAnnoteFile{Stop}{dresden_htz2_2018}%
\bibitem{olsen_z2ht_2019}%
  \BibitemOpen
  \bibfield{author}{%
  \bibinfo {author} {\bibfnamefont{T.}~\bibnamefont{Olsen}}, \bibinfo {author}
  {\bibfnamefont{E.}~\bibnamefont{Andersen}}, \bibinfo {author}
  {\bibfnamefont{T.}~\bibnamefont{Okugawa}}, \bibinfo {author}
  {\bibfnamefont{D.}~\bibnamefont{Torelli}}, \bibinfo {author}
  {\bibfnamefont{T.}~\bibnamefont{Deilmann}},\ and\ \bibinfo {author}
  {\bibfnamefont{K.~S.}\ \bibnamefont{Thygesen}},\ }%
  \Doi{10.1103/PhysRevMaterials.3.024005}{\emph{\bibinfo {title} {Discovering
  Two-Dimensional Topological Insulators from High-Throughput Computations}}},\
  \bibinfo {journal} {Physical Review Materials}\ \textbf{\bibinfo {volume}
  {3}},\ \bibinfo {pages} {024005} (\bibinfo {year} {2019}).~%
  \bibAnnoteFile{Stop}{olsen_z2ht_2019}%
\bibitem{Marrazzo2019}%
  \BibitemOpen
  \bibfield{author}{%
  \bibinfo {author} {\bibfnamefont{A.}~\bibnamefont{Marrazzo}}, \bibinfo
  {author} {\bibfnamefont{M.}~\bibnamefont{Gibertini}}, \bibinfo {author}
  {\bibfnamefont{D.}~\bibnamefont{Campi}}, \bibinfo {author}
  {\bibfnamefont{N.}~\bibnamefont{Mounet}},\ and\ \bibinfo {author}
  {\bibfnamefont{N.}~\bibnamefont{Marzari}},\ }%
  \href{https://arxiv.org/abs/1908.08334}{\emph{\bibinfo {title} {Abundance of
  $\mathbb{Z}_2$ topological order in exfoliable two-dimensional
  insulators}}},\ \bibinfo {journal} {arXiv:1908.08334}\  (\bibinfo {year}
  {2019}).~%
  \bibAnnoteFile{Stop}{Marrazzo2019}%
\bibitem{Marrazzo2018}%
  \BibitemOpen
  \bibfield{author}{%
  \bibinfo {author} {\bibfnamefont{A.}~\bibnamefont{Marrazzo}}, \bibinfo
  {author} {\bibfnamefont{M.}~\bibnamefont{Gibertini}}, \bibinfo {author}
  {\bibfnamefont{D.}~\bibnamefont{Campi}}, \bibinfo {author}
  {\bibfnamefont{N.}~\bibnamefont{Mounet}},\ and\ \bibinfo {author}
  {\bibfnamefont{N.}~\bibnamefont{Marzari}},\ }%
  \Doi{10.1103/PhysRevLett.120.117701}{\emph{\bibinfo {title} {Prediction of a
  Large-Gap and Switchable {Kane-Mele} Quantum Spin {Hall} Insulator}}},\
  \bibinfo {journal} {Phys. Rev. Lett.}\ \textbf{\bibinfo {volume} {120}},\
  \bibinfo {pages} {117701} (\bibinfo {year} {2018}).~%
  \bibAnnoteFile{Stop}{Marrazzo2018}%
\bibitem{facio_prm_2019}%
  \BibitemOpen
  \bibfield{author}{%
  \bibinfo {author} {\bibfnamefont{J.~I.}\ \bibnamefont{Facio}}, \bibinfo
  {author} {\bibfnamefont{S.~K.}\ \bibnamefont{Das}}, \bibinfo {author}
  {\bibfnamefont{Y.}~\bibnamefont{Zhang}}, \bibinfo {author}
  {\bibfnamefont{K.}~\bibnamefont{Koepernik}}, \bibinfo {author}
  {\bibfnamefont{J.}~\bibnamefont{van~den Brink}},\ and\ \bibinfo {author}
  {\bibfnamefont{I.~C.}\ \bibnamefont{Fulga}},\ }%
  \Doi{10.1103/PhysRevMaterials.3.074202}{\emph{\bibinfo {title} {Dual topology
  in jacutingaite ${\mathrm{Pt}}_{2}{\mathrm{HgSe}}_{3}$}}},\ \bibinfo
  {journal} {Phys. Rev. Materials}\ \textbf{\bibinfo {volume} {3}},\ \bibinfo
  {pages} {074202} (\bibinfo {year} {2019}).~%
  \bibAnnoteFile{Stop}{facio_prm_2019}%
\bibitem{bansil_arxiv_2019}%
  \BibitemOpen
  \bibfield{author}{%
  \bibinfo {author} {\bibfnamefont{B.}~\bibnamefont{Ghosh}}, \bibinfo {author}
  {\bibfnamefont{S.}~\bibnamefont{Mardanya}}, \bibinfo {author}
  {\bibfnamefont{B.}~\bibnamefont{Singh}}, \bibinfo {author}
  {\bibfnamefont{X.}~\bibnamefont{Zhou}}, \bibinfo {author}
  {\bibfnamefont{B.}~\bibnamefont{Wang}}, \bibinfo {author}
  {\bibfnamefont{T.-R.}\ \bibnamefont{Chang}}, \bibinfo {author}
  {\bibfnamefont{C.}~\bibnamefont{Su}}, \bibinfo {author}
  {\bibfnamefont{H.}~\bibnamefont{Lin}}, \bibinfo {author}
  {\bibfnamefont{A.}~\bibnamefont{Agarwal}},\ and\ \bibinfo {author}
  {\bibfnamefont{A.}~\bibnamefont{Bansil}},\ }%
  \href{http://arxiv.org/abs/1905.12578}{\emph{\bibinfo {title} {Saddle-point
  von {Hove} singularity and dual topological insulator state in
  {Pt}$_2${HgSe}$_3$}}},\ \bibinfo {journal} {arXiv:1905.12578}\  (\bibinfo
  {year} {2019}).~%
  \bibAnnoteFile{Stop}{bansil_arxiv_2019}%
\bibitem{theory_jacu_2019}%
  \BibitemOpen
  \bibfield{author}{%
  \bibinfo {author} {\bibfnamefont{A.}~\bibnamefont{Marrazzo}}, \bibinfo
  {author} {\bibfnamefont{N.}~\bibnamefont{Marzari}},\ and\ \bibinfo {author}
  {\bibfnamefont{M.}~\bibnamefont{Gibertini}},\ }%
  \href{https://arxiv.org/abs/1909.05050}{\emph{\bibinfo {title} {Emergent dual
  topology in the three-dimensional Kane-Mele
  $\mathrm{Pt}_2\mathrm{HgSe}_3$}}},\ \bibinfo {journal} {arXiv:1909.05050}\
  (\bibinfo {year} {2019}).~%
  \bibAnnoteFile{Stop}{theory_jacu_2019}%
\bibitem{exp_jacu_2019}%
  \BibitemOpen
  \bibfield{author}{%
  \bibinfo {author} {\bibfnamefont{I.}~\bibnamefont{Cucchi}}, \bibinfo {author}
  {\bibfnamefont{A.}~\bibnamefont{Marrazzo}}, \bibinfo {author}
  {\bibfnamefont{E.}~\bibnamefont{Cappelli}}, \bibinfo {author}
  {\bibfnamefont{S.}~\bibnamefont{Ricco}}, \bibinfo {author}
  {\bibfnamefont{F.~Y.}\ \bibnamefont{Bruno}}, \bibinfo {author}
  {\bibfnamefont{S.}~\bibnamefont{Lisi}}, \bibinfo {author}
  {\bibfnamefont{M.}~\bibnamefont{Hoesch}}, \bibinfo {author}
  {\bibfnamefont{T.~K.}\ \bibnamefont{Kim}}, \bibinfo {author}
  {\bibfnamefont{C.}~\bibnamefont{Cacho}}, \bibinfo {author}
  {\bibfnamefont{C.}~\bibnamefont{Besnard}}, \bibinfo {author}
  {\bibfnamefont{E.}~\bibnamefont{Giannini}}, \bibinfo {author}
  {\bibfnamefont{N.}~\bibnamefont{Marzari}}, \bibinfo {author}
  {\bibfnamefont{M.}~\bibnamefont{Gibertini}}, \bibinfo {author}
  {\bibfnamefont{F.}~\bibnamefont{Baumberger}},\ and\ \bibinfo {author}
  {\bibfnamefont{A.}~\bibnamefont{Tamai}},\ }%
  \href{https://arxiv.org/abs/1909.05051}{\emph{\bibinfo {title} {Bulk and
  surface electronic structure of the dual-topology semimetal
  $\mathrm{Pt}_2\mathrm{HgSe}_3$}}},\ \bibinfo {journal} {arXiv:1909.05051}\
  (\bibinfo {year} {2019}).~%
  \bibAnnoteFile{Stop}{exp_jacu_2019}%
\bibitem{Kandrai2019}%
  \BibitemOpen
  \bibfield{author}{%
  \bibinfo {author} {\bibfnamefont{K.}~\bibnamefont{Kandrai}}, \bibinfo
  {author} {\bibfnamefont{G.}~\bibnamefont{Kukucska}}, \bibinfo {author}
  {\bibfnamefont{P.}~\bibnamefont{Vancs{\'o}}}, \bibinfo {author}
  {\bibfnamefont{J.}~\bibnamefont{Koltai}}, \bibinfo {author}
  {\bibfnamefont{G.}~\bibnamefont{Baranka}}, \bibinfo {author}
  {\bibfnamefont{Z.~E.}\ \bibnamefont{Horv{\'a}th}}, \bibinfo {author}
  {\bibfnamefont{{\'A}.}~\bibnamefont{Hoffmann}}, \bibinfo {author}
  {\bibfnamefont{A.}~\bibnamefont{Vymazalov{\'a}}}, \bibinfo {author}
  {\bibfnamefont{L.}~\bibnamefont{Tapaszt{\'o}}},\ and\ \bibinfo {author}
  {\bibfnamefont{P.}~\bibnamefont{Nemes-Incze}},\ }%
  \href{https://arxiv.org/abs/1903.02458}{\emph{\bibinfo {title} {Evidence for
  room temperature quantum spin Hall state in the layered mineral
  jacutingaite}}},\ \bibinfo {journal} {arXiv:1903.02458}\  (\bibinfo {year}
  {2019}).~%
  \bibAnnoteFile{Stop}{Kandrai2019}%
\bibitem{Wu2019}%
  \BibitemOpen
  \bibfield{author}{%
  \bibinfo {author} {\bibfnamefont{X.}~\bibnamefont{Wu}}, \bibinfo {author}
  {\bibfnamefont{M.}~\bibnamefont{Fink}}, \bibinfo {author}
  {\bibfnamefont{W.}~\bibnamefont{Hanke}}, \bibinfo {author}
  {\bibfnamefont{R.}~\bibnamefont{Thomale}},\ and\ \bibinfo {author}
  {\bibfnamefont{D.}~\bibnamefont{Di~Sante}},\ }%
  \Doi{10.1103/PhysRevB.100.041117}{\emph{\bibinfo {title} {Unconventional
  superconductivity in a doped quantum spin Hall insulator}}},\ \bibinfo
  {journal} {Phys. Rev. B}\ \textbf{\bibinfo {volume} {100}},\ \bibinfo {pages}
  {041117(R)} (\bibinfo {year} {2019}).~%
  \bibAnnoteFile{Stop}{Wu2019}%
\bibitem{cabral_first_obs_08}%
  \BibitemOpen
  \bibfield{author}{%
  \bibinfo {author} {\bibfnamefont{A.~R.}\ \bibnamefont{Cabral}}, \bibinfo
  {author} {\bibfnamefont{H.~F.}\ \bibnamefont{Galbiatti}}, \bibinfo {author}
  {\bibfnamefont{R.}~\bibnamefont{Kwitko-Ribeiro}},\ and\ \bibinfo {author}
  {\bibfnamefont{B.}~\bibnamefont{Lehmann}},\ }%
  \Doi{10.1111/j.1365-3121.2007.00783.x}{\emph{\bibinfo {title} {Platinum
  Enrichment at Low Temperatures and Related Microstructures, with Examples of
  Hongshiite ({{PtCu}}) and Empirical `$\mathrm{Pt}_2\mathrm{HgSe}_3$' from
  {Itabira}, {Minas Gerais}, {Brazil}}}},\ \bibinfo {journal} {Terra Nova}\
  \textbf{\bibinfo {volume} {20}},\ \bibinfo {pages} {32} (\bibinfo {year}
  {2008}).~%
  \bibAnnoteFile{Stop}{cabral_first_obs_08}%
\bibitem{Wang2015}%
  \BibitemOpen
  \bibfield{author}{%
  \bibinfo {author} {\bibfnamefont{L.}~\bibnamefont{Wang}}, \bibinfo {author}
  {\bibfnamefont{I.}~\bibnamefont{Guti{\'{e}}rrez-Lezama}}, \bibinfo {author}
  {\bibfnamefont{C.}~\bibnamefont{Barreteau}}, \bibinfo {author}
  {\bibfnamefont{N.}~\bibnamefont{Ubrig}}, \bibinfo {author}
  {\bibfnamefont{E.}~\bibnamefont{Giannini}},\ and\ \bibinfo {author}
  {\bibfnamefont{A.~F.}\ \bibnamefont{Morpurgo}},\ }%
  \Doi{10.1038/ncomms9892}{\emph{\bibinfo {title} {{Tuning magnetotransport in
  a compensated semimetal at the atomic scale}}}},\ \bibinfo {journal} {Nature
  Communications}\ \textbf{\bibinfo {volume} {6}},\ \bibinfo {pages} {1}
  (\bibinfo {year} {2015}).~%
  \bibAnnoteFile{Stop}{Wang2015}%
\bibitem{Wang2016}%
  \BibitemOpen
  \bibfield{author}{%
  \bibinfo {author} {\bibfnamefont{L.}~\bibnamefont{Wang}}, \bibinfo {author}
  {\bibfnamefont{I.}~\bibnamefont{Guti\'errez-Lezama}}, \bibinfo {author}
  {\bibfnamefont{C.}~\bibnamefont{Barreteau}}, \bibinfo {author}
  {\bibfnamefont{D.-K.}\ \bibnamefont{Ki}}, \bibinfo {author}
  {\bibfnamefont{E.}~\bibnamefont{Giannini}},\ and\ \bibinfo {author}
  {\bibfnamefont{A.~F.}\ \bibnamefont{Morpurgo}},\ }%
  \Doi{10.1103/PhysRevLett.117.176601}{\emph{\bibinfo {title} {Direct
  Observation of a Long-Range Field Effect from Gate Tuning of Nonlocal
  Conductivity}}},\ \bibinfo {journal} {Phys. Rev. Lett.}\ \textbf{\bibinfo
  {volume} {117}},\ \bibinfo {pages} {176601} (\bibinfo {year} {2016}).~%
  \bibAnnoteFile{Stop}{Wang2016}%
\bibitem{jacutingaite_exp_12}%
  \BibitemOpen
  \bibfield{author}{%
  \bibinfo {author} {\bibfnamefont{A.}~\bibnamefont{Vymazalov{\'a}}}, \bibinfo
  {author} {\bibfnamefont{F.}~\bibnamefont{Laufek}}, \bibinfo {author}
  {\bibfnamefont{M.}~\bibnamefont{Dr{\'a}bek}}, \bibinfo {author}
  {\bibfnamefont{A.~R.}\ \bibnamefont{Cabral}}, \bibinfo {author}
  {\bibfnamefont{J.}~\bibnamefont{Haloda}}, \bibinfo {author}
  {\bibfnamefont{T.}~\bibnamefont{Sidorinov{\'a}}}, \bibinfo {author}
  {\bibfnamefont{B.}~\bibnamefont{Lehmann}}, \bibinfo {author}
  {\bibfnamefont{H.~F.}\ \bibnamefont{Galbiatti}},\ and\ \bibinfo {author}
  {\bibfnamefont{J.}~\bibnamefont{Drahokoupil}},\ }%
  \Doi{10.3749/canmin.50.2.431}{\emph{\bibinfo {title} {Jacutingaite,
  $\mathrm{Pt}_2\mathrm{HgSe}_3$, a new platinum-group mineral species from the
  Cau{\^e} iron-ore deposit, Itabira district, Minas Gerais, Brazil}}},\
  \bibinfo {journal} {The Canadian Mineralogist}\ \textbf{\bibinfo {volume}
  {50}},\ \bibinfo {pages} {431} (\bibinfo {year} {2012}).~%
  \bibAnnoteFile{Stop}{jacutingaite_exp_12}%
\bibitem{O_Brien_2016}%
  \BibitemOpen
  \bibfield{author}{%
  \bibinfo {author} {\bibfnamefont{M.}~\bibnamefont{O'Brien}}, \bibinfo
  {author} {\bibfnamefont{N.}~\bibnamefont{McEvoy}}, \bibinfo {author}
  {\bibfnamefont{C.}~\bibnamefont{Motta}}, \bibinfo {author}
  {\bibfnamefont{J.-Y.}\ \bibnamefont{Zheng}}, \bibinfo {author}
  {\bibfnamefont{N.~C.}\ \bibnamefont{Berner}}, \bibinfo {author}
  {\bibfnamefont{J.}~\bibnamefont{Kotakoski}}, \bibinfo {author}
  {\bibfnamefont{K.}~\bibnamefont{Elibol}}, \bibinfo {author}
  {\bibfnamefont{T.~J.}\ \bibnamefont{Pennycook}}, \bibinfo {author}
  {\bibfnamefont{J.~C.}\ \bibnamefont{Meyer}}, \bibinfo {author}
  {\bibfnamefont{C.}~\bibnamefont{Yim}}, \bibinfo {author}
  {\bibfnamefont{M.}~\bibnamefont{Abid}}, \bibinfo {author}
  {\bibfnamefont{T.}~\bibnamefont{Hallam}}, \bibinfo {author}
  {\bibfnamefont{J.~F.}\ \bibnamefont{Donegan}}, \bibinfo {author}
  {\bibfnamefont{S.}~\bibnamefont{Sanvito}},\ and\ \bibinfo {author}
  {\bibfnamefont{G.~S.}\ \bibnamefont{Duesberg}},\ }%
  \Doi{10.1088/2053-1583/3/2/021004}{\emph{\bibinfo {title} {Raman
  characterization of platinum diselenide thin films}}},\ \bibinfo {journal}
  {2D Materials}\ \textbf{\bibinfo {volume} {3}},\ \bibinfo {pages} {021004}
  (\bibinfo {year} {2016}).~%
  \bibAnnoteFile{Stop}{O_Brien_2016}%
\bibitem{Xu2011}%
  \BibitemOpen
  \bibfield{author}{%
  \bibinfo {author} {\bibfnamefont{G.}~\bibnamefont{Xu}}, \bibinfo {author}
  {\bibfnamefont{H.}~\bibnamefont{Weng}}, \bibinfo {author}
  {\bibfnamefont{Z.}~\bibnamefont{Wang}}, \bibinfo {author}
  {\bibfnamefont{X.}~\bibnamefont{Dai}},\ and\ \bibinfo {author}
  {\bibfnamefont{Z.}~\bibnamefont{Fang}},\ }%
  \Doi{10.1103/PhysRevLett.107.186806}{\emph{\bibinfo {title} {{Chern semimetal
  and the quantized anomalous Hall effect in HgCr 2 Se 4}}}},\ \bibinfo
  {journal} {Physical Review Letters}\ \textbf{\bibinfo {volume} {107}},\
  \bibinfo {pages} {1} (\bibinfo {year} {2011}).~%
  \bibAnnoteFile{Stop}{Xu2011}%
\bibitem{Liang2015}%
  \BibitemOpen
  \bibfield{author}{%
  \bibinfo {author} {\bibfnamefont{T.}~\bibnamefont{Liang}}, \bibinfo {author}
  {\bibfnamefont{Q.}~\bibnamefont{Gibson}}, \bibinfo {author}
  {\bibfnamefont{M.~N.}\ \bibnamefont{Ali}}, \bibinfo {author}
  {\bibfnamefont{M.}~\bibnamefont{Liu}}, \bibinfo {author}
  {\bibfnamefont{R.~J.}\ \bibnamefont{Cava}},\ and\ \bibinfo {author}
  {\bibfnamefont{N.~P.}\ \bibnamefont{Ong}},\ }%
  \Doi{10.1038/nmat4143}{\emph{\bibinfo {title} {{Ultrahigh mobility and giant
  magnetoresistance in the Dirac semimetal Cd3As2}}}},\ \bibinfo {journal}
  {Nature Materials}\ \textbf{\bibinfo {volume} {14}},\ \bibinfo {pages} {280}
  (\bibinfo {year} {2015}).~%
  \bibAnnoteFile{Stop}{Liang2015}%
\bibitem{Shekhar2015}%
  \BibitemOpen
  \bibfield{author}{%
  \bibinfo {author} {\bibfnamefont{C.}~\bibnamefont{Shekhar}}, \bibinfo
  {author} {\bibfnamefont{A.~K.}\ \bibnamefont{Nayak}}, \bibinfo {author}
  {\bibfnamefont{Y.}~\bibnamefont{Sun}}, \bibinfo {author}
  {\bibfnamefont{M.}~\bibnamefont{Schmidt}}, \bibinfo {author}
  {\bibfnamefont{M.}~\bibnamefont{Nicklas}}, \bibinfo {author}
  {\bibfnamefont{I.}~\bibnamefont{Leermakers}}, \bibinfo {author}
  {\bibfnamefont{U.}~\bibnamefont{Zeitler}}, \bibinfo {author}
  {\bibfnamefont{Y.}~\bibnamefont{Skourski}}, \bibinfo {author}
  {\bibfnamefont{J.}~\bibnamefont{Wosnitza}}, \bibinfo {author}
  {\bibfnamefont{Z.}~\bibnamefont{Liu}}, \bibinfo {author}
  {\bibfnamefont{Y.}~\bibnamefont{Chen}}, \bibinfo {author}
  {\bibfnamefont{W.}~\bibnamefont{Schnelle}}, \bibinfo {author}
  {\bibfnamefont{H.}~\bibnamefont{Borrmann}}, \bibinfo {author}
  {\bibfnamefont{Y.}~\bibnamefont{Grin}}, \bibinfo {author}
  {\bibfnamefont{C.}~\bibnamefont{Felser}},\ and\ \bibinfo {author}
  {\bibfnamefont{B.}~\bibnamefont{Yan}},\ }%
  \Doi{10.1038/nphys3372}{\emph{\bibinfo {title} {{Extremely large
  magnetoresistance and ultrahigh mobility in the topological Weyl semimetal
  candidate NbP}}}},\ \bibinfo {journal} {Nature Physics}\ \textbf{\bibinfo
  {volume} {11}},\ \bibinfo {pages} {645} (\bibinfo {year} {2015}).~%
  \bibAnnoteFile{Stop}{Shekhar2015}%
\bibitem{Leahy10570}%
  \BibitemOpen
  \bibfield{author}{%
  \bibinfo {author} {\bibfnamefont{I.~A.}\ \bibnamefont{Leahy}}, \bibinfo
  {author} {\bibfnamefont{Y.-P.}\ \bibnamefont{Lin}}, \bibinfo {author}
  {\bibfnamefont{P.~E.}\ \bibnamefont{Siegfried}}, \bibinfo {author}
  {\bibfnamefont{A.~C.}\ \bibnamefont{Treglia}}, \bibinfo {author}
  {\bibfnamefont{J.~C.~W.}\ \bibnamefont{Song}}, \bibinfo {author}
  {\bibfnamefont{R.~M.}\ \bibnamefont{Nandkishore}},\ and\ \bibinfo {author}
  {\bibfnamefont{M.}~\bibnamefont{Lee}},\ }%
  \Doi{10.1073/pnas.1808747115}{\emph{\bibinfo {title} {Nonsaturating large
  magnetoresistance in semimetals}}},\ \bibinfo {journal} {Proceedings of the
  National Academy of Sciences}\ \textbf{\bibinfo {volume} {115}},\ \bibinfo
  {pages} {10570} (\bibinfo {year} {2018}).~%
  \bibAnnoteFile{Stop}{Leahy10570}%
\bibitem{ziman1972principles}%
  \BibitemOpen
  \bibfield{author}{%
  \bibinfo {author} {\bibfnamefont{J.}~\bibnamefont{Ziman}},\ }%
  \emph{\bibinfo {title} {Principles of the Theory of Solids}}\ (\bibinfo
  {publisher} {Cambridge University Press},\ \bibinfo {year} {1972})\ ISBN
  \bibinfo {isbn} {9780521297332}.~%
  \bibAnnoteFile{Stop}{ziman1972principles}%
\bibitem{Schoenberg}%
  \BibitemOpen
  \bibfield{author}{%
  \bibinfo {author} {\bibfnamefont{D.}~\bibnamefont{{Schoenberg}}},\ }%
  \emph{\bibinfo {title} {{Magnetic Oscillations in Metals}}}\ (\bibinfo
  {publisher} {Cambridge University Press},\ \bibinfo {address} {Cambridge},\
  \bibinfo {year} {2009}).~%
  \bibAnnoteFile{Stop}{Schoenberg}%
\bibitem{LK}%
  \BibitemOpen
  \bibfield{author}{%
  \bibinfo {author} {\bibfnamefont{I.}~\bibnamefont{Lifshitz}}\ and\ \bibinfo
  {author} {\bibfnamefont{A.}~\bibnamefont{Kosevich}},\ }%
  \href{http://www.jetp.ac.ru/cgi-bin/e/index/e/2/4/p636?a=list}{\emph{\bibinfo
  {title} {Theory of Magnetic Susceptibility in Metals at Low Temperature}}},\
  \bibinfo {journal} {Sov. Phys. JETP}\ \textbf{\bibinfo {volume} {2}},\
  \bibinfo {pages} {636} (\bibinfo {year} {1956}).~%
  \bibAnnoteFile{Stop}{LK}%
\bibitem{Richards1973}%
  \BibitemOpen
  \bibfield{author}{%
  \bibinfo {author} {\bibfnamefont{F.~E.}\ \bibnamefont{Richards}},\ }%
  \Doi{10.1103/PhysRevB.8.2552}{\emph{\bibinfo {title} {{Investigation of the
  magnetoresistance quantum oscillations in magnesium}}}},\ \bibinfo {journal}
  {Physical Review B}\ \textbf{\bibinfo {volume} {8}},\ \bibinfo {pages} {2552}
  (\bibinfo {year} {1973}).~%
  \bibAnnoteFile{Stop}{Richards1973}%
\bibitem{Niederer1974}%
  \BibitemOpen
  \bibfield{author}{%
  \bibinfo {author} {\bibfnamefont{H.~H.}\ \bibnamefont{Niederer}},\ }%
  \Doi{10.7567/JJAPS.2S2.339}{\emph{\bibinfo {title} {{Magneto oscillatory
  conductance in n-type inverted silicon surfaces}}}},\ \bibinfo {journal}
  {Japanese Journal of Applied Physics}\ \textbf{\bibinfo {volume} {13}},\
  \bibinfo {pages} {339} (\bibinfo {year} {1974}).~%
  \bibAnnoteFile{Stop}{Niederer1974}%
\bibitem{Fang1977}%
  \BibitemOpen
  \bibfield{author}{%
  \bibinfo {author} {\bibfnamefont{F.~F.}\ \bibnamefont{Fang}}, \bibinfo
  {author} {\bibfnamefont{A.~B.}\ \bibnamefont{Fowler}},\ and\ \bibinfo
  {author} {\bibfnamefont{A.}~\bibnamefont{Hartstein}},\ }%
  \Doi{10.1103/PhysRevB.16.4446}{\emph{\bibinfo {title} {Effective mass and
  collision time of (100) Si surface electrons}}},\ \bibinfo {journal} {Phys.
  Rev. B}\ \textbf{\bibinfo {volume} {16}},\ \bibinfo {pages} {4446} (\bibinfo
  {year} {1977}).~%
  \bibAnnoteFile{Stop}{Fang1977}%
\bibitem{Balicas2000}%
  \BibitemOpen
  \bibfield{author}{%
  \bibinfo {author} {\bibfnamefont{L.}~\bibnamefont{Balicas}}, \bibinfo
  {author} {\bibfnamefont{J.~S.}\ \bibnamefont{Brooks}}, \bibinfo {author}
  {\bibfnamefont{K.}~\bibnamefont{Storr}}, \bibinfo {author}
  {\bibfnamefont{D.}~\bibnamefont{Graf}}, \bibinfo {author}
  {\bibfnamefont{S.}~\bibnamefont{Uji}}, \bibinfo {author}
  {\bibfnamefont{H.}~\bibnamefont{Shinagawa}}, \bibinfo {author}
  {\bibfnamefont{E.}~\bibnamefont{Ojima}}, \bibinfo {author}
  {\bibfnamefont{H.}~\bibnamefont{Fujiwara}}, \bibinfo {author}
  {\bibfnamefont{H.}~\bibnamefont{Kobayashi}}, \bibinfo {author}
  {\bibfnamefont{A.}~\bibnamefont{Kobayashi}},\ and\ \bibinfo {author}
  {\bibfnamefont{M.}~\bibnamefont{Tokumoto}},\ }%
  \Doi{10.1016/S0038-1098(00)00374-4}{\emph{\bibinfo {title} {{Shubnikov-de
  Haas effect and Yamaji oscillations in the antiferromagnetically ordered
  organic superconductor $\kappa$-(BETS)2FeBr4: A fermiology study}}}},\
  \bibinfo {journal} {Solid State Communications}\ \textbf{\bibinfo {volume}
  {116}},\ \bibinfo {pages} {557} (\bibinfo {year} {2000}).~%
  \bibAnnoteFile{Stop}{Balicas2000}%
\bibitem{Bangura2008}%
  \BibitemOpen
  \bibfield{author}{%
  \bibinfo {author} {\bibfnamefont{A.~F.}\ \bibnamefont{Bangura}}, \bibinfo
  {author} {\bibfnamefont{J.~D.}\ \bibnamefont{Fletcher}}, \bibinfo {author}
  {\bibfnamefont{A.}~\bibnamefont{Carrington}}, \bibinfo {author}
  {\bibfnamefont{J.}~\bibnamefont{Levallois}}, \bibinfo {author}
  {\bibfnamefont{M.}~\bibnamefont{Nardone}}, \bibinfo {author}
  {\bibfnamefont{B.}~\bibnamefont{Vignolle}}, \bibinfo {author}
  {\bibfnamefont{P.~J.}\ \bibnamefont{Heard}}, \bibinfo {author}
  {\bibfnamefont{N.}~\bibnamefont{Doiron-Leyraud}}, \bibinfo {author}
  {\bibfnamefont{D.}~\bibnamefont{Leboeuf}}, \bibinfo {author}
  {\bibfnamefont{L.}~\bibnamefont{Taillefer}}, \bibinfo {author}
  {\bibfnamefont{S.}~\bibnamefont{Adachi}}, \bibinfo {author}
  {\bibfnamefont{C.}~\bibnamefont{Proust}},\ and\ \bibinfo {author}
  {\bibfnamefont{N.~E.}\ \bibnamefont{Hussey}},\ }%
  \Doi{10.1103/PhysRevLett.100.047004}{\emph{\bibinfo {title} {{Small fermi
  surface pockets in underdoped high temperature superconductors: Observation
  of Shubnikov-de Haas oscillations in YBa2Cu4O8}}}},\ \bibinfo {journal}
  {Physical Review Letters}\ \textbf{\bibinfo {volume} {100}},\ \bibinfo
  {pages} {1} (\bibinfo {year} {2008}).~%
  \bibAnnoteFile{Stop}{Bangura2008}%
\bibitem{Cai2015}%
  \BibitemOpen
  \bibfield{author}{%
  \bibinfo {author} {\bibfnamefont{P.~L.}\ \bibnamefont{Cai}}, \bibinfo
  {author} {\bibfnamefont{J.}~\bibnamefont{Hu}}, \bibinfo {author}
  {\bibfnamefont{L.~P.}\ \bibnamefont{He}}, \bibinfo {author}
  {\bibfnamefont{J.}~\bibnamefont{Pan}}, \bibinfo {author}
  {\bibfnamefont{X.~C.}\ \bibnamefont{Hong}}, \bibinfo {author}
  {\bibfnamefont{Z.}~\bibnamefont{Zhang}}, \bibinfo {author}
  {\bibfnamefont{J.}~\bibnamefont{Zhang}}, \bibinfo {author}
  {\bibfnamefont{J.}~\bibnamefont{Wei}}, \bibinfo {author}
  {\bibfnamefont{Z.~Q.}\ \bibnamefont{Mao}},\ and\ \bibinfo {author}
  {\bibfnamefont{S.~Y.}\ \bibnamefont{Li}},\ }%
  \Doi{10.1103/PhysRevLett.115.057202}{\emph{\bibinfo {title} {{Drastic
  Pressure Effect on the Extremely Large Magnetoresistance in WTe2: Quantum
  Oscillation Study}}}},\ \bibinfo {journal} {Physical Review Letters}\
  \textbf{\bibinfo {volume} {115}},\ \bibinfo {pages} {1} (\bibinfo {year}
  {2015}).~%
  \bibAnnoteFile{Stop}{Cai2015}%
\bibitem{Rhodes2015}%
  \BibitemOpen
  \bibfield{author}{%
  \bibinfo {author} {\bibfnamefont{D.}~\bibnamefont{Rhodes}}, \bibinfo {author}
  {\bibfnamefont{S.}~\bibnamefont{Das}}, \bibinfo {author}
  {\bibfnamefont{Q.~R.}\ \bibnamefont{Zhang}}, \bibinfo {author}
  {\bibfnamefont{B.}~\bibnamefont{Zeng}}, \bibinfo {author}
  {\bibfnamefont{N.~R.}\ \bibnamefont{Pradhan}}, \bibinfo {author}
  {\bibfnamefont{N.}~\bibnamefont{Kikugawa}}, \bibinfo {author}
  {\bibfnamefont{E.}~\bibnamefont{Manousakis}},\ and\ \bibinfo {author}
  {\bibfnamefont{L.}~\bibnamefont{Balicas}},\ }%
  \Doi{10.1103/PhysRevB.92.125152}{\emph{\bibinfo {title} {{Role of spin-orbit
  coupling and evolution of the electronic structure of WTe2 under an external
  magnetic field}}}},\ \bibinfo {journal} {Physical Review B - Condensed Matter
  and Materials Physics}\ \textbf{\bibinfo {volume} {92}},\ \bibinfo {pages}
  {1} (\bibinfo {year} {2015}).~%
  \bibAnnoteFile{Stop}{Rhodes2015}%
\bibitem{Note1}%
  \BibitemOpen
  \bibinfo {note} {The experimental spectra used in the analyses are obtained
  by calculating the FT of the the derivative of the SdH oscillations, for
  consistency we use the FT spectra of the derivative of equation \protect
  \textup {\hbox {\mathsurround \z@ \protect \normalfont (\ignorespaces \ref
  {eqn:eq1}\unskip \@@italiccorr )}}}~%
  \bibAnnoteFile{NoStop}{Note1}%
\bibitem{Dion2004}%
  \BibitemOpen
  \bibfield{author}{%
  \bibinfo {author} {\bibfnamefont{M.}~\bibnamefont{Dion}}, \bibinfo {author}
  {\bibfnamefont{H.}~\bibnamefont{Rydberg}}, \bibinfo {author}
  {\bibfnamefont{E.}~\bibnamefont{Schr\"oder}}, \bibinfo {author}
  {\bibfnamefont{D.~C.}\ \bibnamefont{Langreth}},\ and\ \bibinfo {author}
  {\bibfnamefont{B.~I.}\ \bibnamefont{Lundqvist}},\ }%
  \Doi{10.1103/PhysRevLett.92.246401}{\emph{\bibinfo {title} {Van der Waals
  Density Functional for General Geometries}}},\ \bibinfo {journal} {Phys. Rev.
  Lett.}\ \textbf{\bibinfo {volume} {92}},\ \bibinfo {pages} {246401} (\bibinfo
  {year} {2004}).~%
  \bibAnnoteFile{Stop}{Dion2004}%
\bibitem{Berland2014}%
  \BibitemOpen
  \bibfield{author}{%
  \bibinfo {author} {\bibfnamefont{K.}~\bibnamefont{Berland}}\ and\ \bibinfo
  {author} {\bibfnamefont{P.}~\bibnamefont{Hyldgaard}},\ }%
  \Doi{10.1103/PhysRevB.89.035412}{\emph{\bibinfo {title} {Exchange functional
  that tests the robustness of the plasmon description of the van der Waals
  density functional}}},\ \bibinfo {journal} {Phys. Rev. B}\ \textbf{\bibinfo
  {volume} {89}},\ \bibinfo {pages} {035412} (\bibinfo {year} {2014}).~%
  \bibAnnoteFile{Stop}{Berland2014}%
\bibitem{Vydrov2010}%
  \BibitemOpen
  \bibfield{author}{%
  \bibinfo {author} {\bibfnamefont{O.~A.}\ \bibnamefont{Vydrov}}\ and\ \bibinfo
  {author} {\bibfnamefont{T.}~\bibnamefont{Van~Voorhis}},\ }%
  \Doi{10.1063/1.3521275}{\emph{\bibinfo {title} {Nonlocal van der Waals
  density functional: The simpler the better}}},\ \bibinfo {journal} {The
  Journal of Chemical Physics}\ \textbf{\bibinfo {volume} {133}},\ \bibinfo
  {pages} {244103} (\bibinfo {year} {2010}).~%
  \bibAnnoteFile{Stop}{Vydrov2010}%
\bibitem{Sabatini2013}%
  \BibitemOpen
  \bibfield{author}{%
  \bibinfo {author} {\bibfnamefont{R.}~\bibnamefont{Sabatini}}, \bibinfo
  {author} {\bibfnamefont{T.}~\bibnamefont{Gorni}},\ and\ \bibinfo {author}
  {\bibfnamefont{S.}~\bibnamefont{de~Gironcoli}},\ }%
  \Doi{10.1103/PhysRevB.87.041108}{\emph{\bibinfo {title} {Nonlocal van der
  Waals density functional made simple and efficient}}},\ \bibinfo {journal}
  {Phys. Rev. B}\ \textbf{\bibinfo {volume} {87}},\ \bibinfo {pages} {041108}
  (\bibinfo {year} {2013}).~%
  \bibAnnoteFile{Stop}{Sabatini2013}%
\bibitem{giannozzi_quantum_2009}%
  \BibitemOpen
  \bibfield{author}{%
  \bibinfo {author} {\bibfnamefont{P.}~\bibnamefont{Giannozzi}}, \bibinfo
  {author} {\bibfnamefont{S.}~\bibnamefont{Baroni}}, \bibinfo {author}
  {\bibfnamefont{N.}~\bibnamefont{Bonini}}, \bibinfo {author}
  {\bibfnamefont{M.}~\bibnamefont{Calandra}}, \bibinfo {author}
  {\bibfnamefont{R.}~\bibnamefont{Car}}, \bibinfo {author}
  {\bibfnamefont{C.}~\bibnamefont{Cavazzoni}}, \bibinfo {author}
  {\bibnamefont{{Davide Ceresoli}}}, \bibinfo {author} {\bibfnamefont{G.~L.}\
  \bibnamefont{Chiarotti}}, \bibinfo {author}
  {\bibfnamefont{M.}~\bibnamefont{Cococcioni}}, \bibinfo {author}
  {\bibfnamefont{I.}~\bibnamefont{Dabo}}, \bibinfo {author}
  {\bibfnamefont{A.~D.}\ \bibnamefont{Corso}}, \bibinfo {author}
  {\bibfnamefont{S.}~\bibnamefont{de~Gironcoli}}, \bibinfo {author}
  {\bibfnamefont{S.}~\bibnamefont{Fabris}}, \bibinfo {author}
  {\bibfnamefont{G.}~\bibnamefont{Fratesi}}, \bibinfo {author}
  {\bibfnamefont{R.}~\bibnamefont{Gebauer}}, \bibinfo {author}
  {\bibfnamefont{U.}~\bibnamefont{Gerstmann}}, \bibinfo {author}
  {\bibfnamefont{C.}~\bibnamefont{Gougoussis}}, \bibinfo {author}
  {\bibnamefont{{Anton Kokalj}}}, \bibinfo {author}
  {\bibfnamefont{M.}~\bibnamefont{Lazzeri}}, \bibinfo {author}
  {\bibfnamefont{L.}~\bibnamefont{Martin-Samos}}, \bibinfo {author}
  {\bibfnamefont{N.}~\bibnamefont{Marzari}}, \bibinfo {author}
  {\bibfnamefont{F.}~\bibnamefont{Mauri}}, \bibinfo {author}
  {\bibfnamefont{R.}~\bibnamefont{Mazzarello}}, \bibinfo {author}
  {\bibnamefont{{Stefano Paolini}}}, \bibinfo {author}
  {\bibfnamefont{A.}~\bibnamefont{Pasquarello}}, \bibinfo {author}
  {\bibfnamefont{L.}~\bibnamefont{Paulatto}}, \bibinfo {author}
  {\bibfnamefont{C.}~\bibnamefont{Sbraccia}}, \bibinfo {author}
  {\bibfnamefont{S.}~\bibnamefont{Scandolo}}, \bibinfo {author}
  {\bibfnamefont{G.}~\bibnamefont{Sclauzero}}, \bibinfo {author}
  {\bibfnamefont{A.~P.}\ \bibnamefont{Seitsonen}}, \bibinfo {author}
  {\bibfnamefont{A.}~\bibnamefont{Smogunov}}, \bibinfo {author}
  {\bibfnamefont{P.}~\bibnamefont{Umari}},\ and\ \bibinfo {author}
  {\bibfnamefont{R.~M.}\ \bibnamefont{Wentzcovitch}},\ }%
  \Doi{10.1088/0953-8984/21/39/395502}{\emph{\bibinfo {title} {{{QUANTUM
  ESPRESSO}}: A Modular and Open-Source Software Project for Quantum
  Simulations of Materials}}},\ \bibinfo {journal} {Journal of Physics:
  Condensed Matter}\ \textbf{\bibinfo {volume} {21}},\ \bibinfo {pages}
  {395502} (\bibinfo {year} {2009}).~%
  \bibAnnoteFile{Stop}{giannozzi_quantum_2009}%
\bibitem{giannozzi_qe_2017}%
  \BibitemOpen
  \bibfield{author}{%
  \bibinfo {author} {\bibfnamefont{P.}~\bibnamefont{Giannozzi}}, \bibinfo
  {author} {\bibfnamefont{O.}~\bibnamefont{Andreussi}}, \bibinfo {author}
  {\bibfnamefont{T.}~\bibnamefont{Brumme}}, \bibinfo {author}
  {\bibfnamefont{O.}~\bibnamefont{Bunau}}, \bibinfo {author}
  {\bibfnamefont{M.~B.}\ \bibnamefont{Nardelli}}, \bibinfo {author}
  {\bibfnamefont{M.}~\bibnamefont{Calandra}}, \bibinfo {author}
  {\bibfnamefont{R.}~\bibnamefont{Car}}, \bibinfo {author}
  {\bibfnamefont{C.}~\bibnamefont{Cavazzoni}}, \bibinfo {author}
  {\bibnamefont{{D Ceresoli}}}, \bibinfo {author}
  {\bibfnamefont{M.}~\bibnamefont{Cococcioni}}, \bibinfo {author}
  {\bibfnamefont{N.}~\bibnamefont{Colonna}}, \bibinfo {author}
  {\bibfnamefont{I.}~\bibnamefont{Carnimeo}}, \bibinfo {author}
  {\bibfnamefont{A.~D.}\ \bibnamefont{Corso}}, \bibinfo {author}
  {\bibfnamefont{S.}~\bibnamefont{de~Gironcoli}}, \bibinfo {author}
  {\bibfnamefont{P.}~\bibnamefont{Delugas}}, \bibinfo {author}
  {\bibfnamefont{R.~A.~D.}\ \bibnamefont{Jr}}, \bibinfo {author}
  {\bibnamefont{{A Ferretti}}}, \bibinfo {author}
  {\bibfnamefont{A.}~\bibnamefont{Floris}}, \bibinfo {author}
  {\bibfnamefont{G.}~\bibnamefont{Fratesi}}, \bibinfo {author}
  {\bibfnamefont{G.}~\bibnamefont{Fugallo}}, \bibinfo {author}
  {\bibfnamefont{R.}~\bibnamefont{Gebauer}}, \bibinfo {author}
  {\bibfnamefont{U.}~\bibnamefont{Gerstmann}}, \bibinfo {author}
  {\bibfnamefont{F.}~\bibnamefont{Giustino}}, \bibinfo {author}
  {\bibfnamefont{T.}~\bibnamefont{Gorni}}, \bibinfo {author}
  {\bibfnamefont{J.}~\bibnamefont{Jia}}, \bibinfo {author}
  {\bibfnamefont{M.}~\bibnamefont{Kawamura}}, \bibinfo {author}
  {\bibnamefont{{H-Y Ko}}}, \bibinfo {author}
  {\bibfnamefont{A.}~\bibnamefont{Kokalj}}, \bibinfo {author}
  {\bibfnamefont{E.}~\bibnamefont{K{\"u}{\c c}{\"u}kbenli}}, \bibinfo {author}
  {\bibfnamefont{M.}~\bibnamefont{Lazzeri}}, \bibinfo {author}
  {\bibfnamefont{M.}~\bibnamefont{Marsili}}, \bibinfo {author}
  {\bibfnamefont{N.}~\bibnamefont{Marzari}}, \bibinfo {author}
  {\bibfnamefont{F.}~\bibnamefont{Mauri}}, \bibinfo {author}
  {\bibfnamefont{N.~L.}\ \bibnamefont{Nguyen}}, \bibinfo {author}
  {\bibfnamefont{H.-V.}\ \bibnamefont{Nguyen}}, \bibinfo {author}
  {\bibnamefont{{A Otero-de-la-Roza}}}, \bibinfo {author}
  {\bibfnamefont{L.}~\bibnamefont{Paulatto}}, \bibinfo {author}
  {\bibfnamefont{S.}~\bibnamefont{Ponc{\'e}}}, \bibinfo {author}
  {\bibfnamefont{D.}~\bibnamefont{Rocca}}, \bibinfo {author}
  {\bibfnamefont{R.}~\bibnamefont{Sabatini}}, \bibinfo {author}
  {\bibfnamefont{B.}~\bibnamefont{Santra}}, \bibinfo {author}
  {\bibfnamefont{M.}~\bibnamefont{Schlipf}}, \bibinfo {author}
  {\bibfnamefont{A.~P.}\ \bibnamefont{Seitsonen}}, \bibinfo {author}
  {\bibfnamefont{A.}~\bibnamefont{Smogunov}}, \bibinfo {author}
  {\bibnamefont{{I Timrov}}}, \bibinfo {author}
  {\bibfnamefont{T.}~\bibnamefont{Thonhauser}}, \bibinfo {author}
  {\bibfnamefont{P.}~\bibnamefont{Umari}}, \bibinfo {author}
  {\bibfnamefont{N.}~\bibnamefont{Vast}}, \bibinfo {author}
  {\bibfnamefont{X.}~\bibnamefont{Wu}},\ and\ \bibinfo {author}
  {\bibfnamefont{S.}~\bibnamefont{Baroni}},\ }%
  \Doi{10.1088/1361-648X/aa8f79}{\emph{\bibinfo {title} {Advanced Capabilities
  for Materials Modelling with {{Q}} Uantum {{ESPRESSO}}}}},\ \bibinfo
  {journal} {Journal of Physics: Condensed Matter}\ \textbf{\bibinfo {volume}
  {29}},\ \bibinfo {pages} {465901} (\bibinfo {year} {2017}).~%
  \bibAnnoteFile{Stop}{giannozzi_qe_2017}%
\bibitem{Onsager1952}%
  \BibitemOpen
  \bibfield{author}{%
  \bibinfo {author} {\bibfnamefont{L.}~\bibnamefont{Onsager}},\ }%
  \Doi{10.1080/14786440908521019}{\emph{\bibinfo {title} {Interpretation of the
  de Haas-van Alphen effect}}},\ \bibinfo {journal} {Phil. Mag.}\
  \textbf{\bibinfo {volume} {43}},\ \bibinfo {pages} {1006} (\bibinfo {year}
  {1952}).~%
  \bibAnnoteFile{Stop}{Onsager1952}%
\bibitem{Drabek2012}%
  \BibitemOpen
  \bibfield{author}{%
  \bibinfo {author} {\bibfnamefont{M.}~\bibnamefont{Dr{\'{a}}bek}}, \bibinfo
  {author} {\bibfnamefont{A.}~\bibnamefont{Vymazalov{\'{a}}}},\ and\ \bibinfo
  {author} {\bibfnamefont{A.~R.}\ \bibnamefont{Cabral}},\ }%
  \Doi{10.3749/canmin.50.2.441}{\emph{\bibinfo {title} {{The system Hg-Pt-Se At
  400$\,^{\circ}$C: Phase relations involving jacutingaite}}}},\ \bibinfo
  {journal} {Canadian Mineralogist}\ \textbf{\bibinfo {volume} {50}},\ \bibinfo
  {pages} {441} (\bibinfo {year} {2012}).~%
  \bibAnnoteFile{Stop}{Drabek2012}%
\bibitem{prandini_precision_2018}%
  \BibitemOpen
  \bibfield{author}{%
  \bibinfo {author} {\bibfnamefont{G.}~\bibnamefont{Prandini}}, \bibinfo
  {author} {\bibfnamefont{A.}~\bibnamefont{Marrazzo}}, \bibinfo {author}
  {\bibfnamefont{I.~E.}\ \bibnamefont{Castelli}}, \bibinfo {author}
  {\bibfnamefont{N.}~\bibnamefont{Mounet}},\ and\ \bibinfo {author}
  {\bibfnamefont{N.}~\bibnamefont{Marzari}},\ }%
  \Doi{10.1038/s41524-018-0127-2}{\emph{\bibinfo {title} {Precision and
  efficiency in solid-state pseudopotential calculations}}},\ \bibinfo
  {journal} {npj Computational Materials}\ \textbf{\bibinfo {volume} {4}},\
  \bibinfo {pages} {72} (\bibinfo {year} {2018}).~%
  \bibAnnoteFile{Stop}{prandini_precision_2018}%
\bibitem{dojo_paper_18}%
  \BibitemOpen
  \bibfield{author}{%
  \bibinfo {author} {\bibfnamefont{M.~J.}\ \bibnamefont{van Setten}}, \bibinfo
  {author} {\bibfnamefont{M.}~\bibnamefont{Giantomassi}}, \bibinfo {author}
  {\bibfnamefont{E.}~\bibnamefont{Bousquet}}, \bibinfo {author}
  {\bibfnamefont{M.~J.}\ \bibnamefont{Verstraete}}, \bibinfo {author}
  {\bibfnamefont{D.~R.}\ \bibnamefont{Hamann}}, \bibinfo {author}
  {\bibfnamefont{X.}~\bibnamefont{Gonze}},\ and\ \bibinfo {author}
  {\bibfnamefont{G.~M.}\ \bibnamefont{Rignanese}},\ }%
  \Doi{10.1016/j.cpc.2018.01.012}{\emph{\bibinfo {title} {The {PseudoDojo}:
  {Training} and grading a 85 element optimized norm-conserving pseudopotential
  table}}},\ \bibinfo {journal} {Computer Physics Communications}\
  \textbf{\bibinfo {volume} {226}},\ \bibinfo {pages} {39} (\bibinfo {year}
  {2018}).~%
  \bibAnnoteFile{Stop}{dojo_paper_18}%
\bibitem{mv_smearing_99}%
  \BibitemOpen
  \bibfield{author}{%
  \bibinfo {author} {\bibfnamefont{N.}~\bibnamefont{Marzari}}, \bibinfo
  {author} {\bibfnamefont{D.}~\bibnamefont{Vanderbilt}}, \bibinfo {author}
  {\bibfnamefont{A.}~\bibnamefont{De~Vita}},\ and\ \bibinfo {author}
  {\bibfnamefont{M.~C.}\ \bibnamefont{Payne}},\ }%
  \Doi{10.1103/PhysRevLett.82.3296}{\emph{\bibinfo {title} {Thermal Contraction
  and Disordering of the Al(110) Surface}}},\ \bibinfo {journal} {Phys. Rev.
  Lett.}\ \textbf{\bibinfo {volume} {82}},\ \bibinfo {pages} {3296} (\bibinfo
  {year} {1999}).~%
  \bibAnnoteFile{Stop}{mv_smearing_99}%
\bibitem{wannier_review_12}%
  \BibitemOpen
  \bibfield{author}{%
  \bibinfo {author} {\bibfnamefont{N.}~\bibnamefont{Marzari}}, \bibinfo
  {author} {\bibfnamefont{A.~A.}\ \bibnamefont{Mostofi}}, \bibinfo {author}
  {\bibfnamefont{J.~R.}\ \bibnamefont{Yates}}, \bibinfo {author}
  {\bibfnamefont{I.}~\bibnamefont{Souza}},\ and\ \bibinfo {author}
  {\bibfnamefont{D.}~\bibnamefont{Vanderbilt}},\ }%
  \Doi{10.1103/RevModPhys.84.1419}{\emph{\bibinfo {title} {Maximally Localized
  {{Wannier}} Functions: {{Theory}} and Applications}}},\ \bibinfo {journal}
  {Reviews of Modern Physics}\ \textbf{\bibinfo {volume} {84}},\ \bibinfo
  {pages} {1419} (\bibinfo {year} {2012}).~%
  \bibAnnoteFile{Stop}{wannier_review_12}%
\bibitem{mostofi_updated_2014}%
  \BibitemOpen
  \bibfield{author}{%
  \bibinfo {author} {\bibfnamefont{A.~A.}\ \bibnamefont{Mostofi}}, \bibinfo
  {author} {\bibfnamefont{J.~R.}\ \bibnamefont{Yates}}, \bibinfo {author}
  {\bibfnamefont{G.}~\bibnamefont{Pizzi}}, \bibinfo {author}
  {\bibfnamefont{Y.-S.}\ \bibnamefont{Lee}}, \bibinfo {author}
  {\bibfnamefont{I.}~\bibnamefont{Souza}}, \bibinfo {author}
  {\bibfnamefont{D.}~\bibnamefont{Vanderbilt}},\ and\ \bibinfo {author}
  {\bibfnamefont{N.}~\bibnamefont{Marzari}},\ }%
  \Doi{10.1016/j.cpc.2014.05.003}{\emph{\bibinfo {title} {An Updated Version of
  Wannier90: {{A}} Tool for Obtaining Maximally-Localised {{Wannier}}
  Functions}}},\ \bibinfo {journal} {Computer Physics Communications}\
  \textbf{\bibinfo {volume} {185}},\ \bibinfo {pages} {2309} (\bibinfo {year}
  {2014}).~%
  \bibAnnoteFile{Stop}{mostofi_updated_2014}%
\bibitem{Rourke2012}%
  \BibitemOpen
  \bibfield{author}{%
  \bibinfo {author} {\bibfnamefont{P.}~\bibnamefont{Rourke}}\ and\ \bibinfo
  {author} {\bibfnamefont{S.}~\bibnamefont{Julian}},\ }%
  \Doi{https://doi.org/10.1016/j.cpc.2011.10.015}{\emph{\bibinfo {title}
  {Numerical extraction of de Haas--van Alphen frequencies from calculated band
  energies}}},\ \bibinfo {journal} {Computer Physics Communications}\
  \textbf{\bibinfo {volume} {183}},\ \bibinfo {pages} {324 } (\bibinfo {year}
  {2012}).~%
  \bibAnnoteFile{Stop}{Rourke2012}%
\bibitem{wannier_tools_18}%
  \BibitemOpen
  \bibfield{author}{%
  \bibinfo {author} {\bibfnamefont{Q.}~\bibnamefont{Wu}}, \bibinfo {author}
  {\bibfnamefont{S.}~\bibnamefont{Zhang}}, \bibinfo {author}
  {\bibfnamefont{H.-F.}\ \bibnamefont{Song}}, \bibinfo {author}
  {\bibfnamefont{M.}~\bibnamefont{Troyer}},\ and\ \bibinfo {author}
  {\bibfnamefont{A.~A.}\ \bibnamefont{Soluyanov}},\ }%
  \Doi{https://doi.org/10.1016/j.cpc.2017.09.033}{\emph{\bibinfo {title}
  {{WannierTools}: {An} open-source software package for novel topological
  materials}}},\ \bibinfo {journal} {Computer Physics Communications}\
  \textbf{\bibinfo {volume} {224}},\ \bibinfo {pages} {405 } (\bibinfo {year}
  {2018}).~%
  \bibAnnoteFile{Stop}{wannier_tools_18}%
\bibitem{Lee2010}%
  \BibitemOpen
  \bibfield{author}{%
  \bibinfo {author} {\bibfnamefont{K.}~\bibnamefont{Lee}}, \bibinfo {author}
  {\bibfnamefont{E.~D.}\ \bibnamefont{Murray}}, \bibinfo {author}
  {\bibfnamefont{L.}~\bibnamefont{Kong}}, \bibinfo {author}
  {\bibfnamefont{B.~I.}\ \bibnamefont{Lundqvist}},\ and\ \bibinfo {author}
  {\bibfnamefont{D.~C.}\ \bibnamefont{Langreth}},\ }%
  \Doi{10.1103/PhysRevB.82.081101}{\emph{\bibinfo {title} {Higher-accuracy van
  der Waals density functional}}},\ \bibinfo {journal} {Phys. Rev. B}\
  \textbf{\bibinfo {volume} {82}},\ \bibinfo {pages} {081101} (\bibinfo {year}
  {2010}).~%
  \bibAnnoteFile{Stop}{Lee2010}%
\bibitem{phonopy}%
  \BibitemOpen
  \bibfield{author}{%
  \bibinfo {author} {\bibfnamefont{A.}~\bibnamefont{Togo}}\ and\ \bibinfo
  {author} {\bibfnamefont{I.}~\bibnamefont{Tanaka}},\ }%
  \emph{\bibinfo {title} {First principles phonon calculations in materials
  science}},\ \bibinfo {journal} {Scr. Mater.}\ \textbf{\bibinfo {volume}
  {108}},\ \bibinfo {pages} {1} (\bibinfo {year} {2015}).~%
  \bibAnnoteFile{Stop}{phonopy}%
\bibitem{Cooper2010}%
  \BibitemOpen
  \bibfield{author}{%
  \bibinfo {author} {\bibfnamefont{V.~R.}\ \bibnamefont{Cooper}},\ }%
  \Doi{10.1103/PhysRevB.81.161104}{\emph{\bibinfo {title} {Van der Waals
  density functional: An appropriate exchange functional}}},\ \bibinfo
  {journal} {Phys. Rev. B}\ \textbf{\bibinfo {volume} {81}},\ \bibinfo {pages}
  {161104} (\bibinfo {year} {2010}).~%
  \bibAnnoteFile{Stop}{Cooper2010}%
\bibitem{Klimes2011}%
  \BibitemOpen
  \bibfield{author}{%
  \bibinfo {author} {\bibfnamefont{J.~c.~v.}\
  \bibnamefont{Klime\ifmmode~\check{s}\else \v{s}\fi{}}}, \bibinfo {author}
  {\bibfnamefont{D.~R.}\ \bibnamefont{Bowler}},\ and\ \bibinfo {author}
  {\bibfnamefont{A.}~\bibnamefont{Michaelides}},\ }%
  \Doi{10.1103/PhysRevB.83.195131}{\emph{\bibinfo {title} {Van der Waals
  density functionals applied to solids}}},\ \bibinfo {journal} {Phys. Rev. B}\
  \textbf{\bibinfo {volume} {83}},\ \bibinfo {pages} {195131} (\bibinfo {year}
  {2011}).~%
  \bibAnnoteFile{Stop}{Klimes2011}%
\bibitem{Klimes2009}%
  \BibitemOpen
  \bibfield{author}{%
  \bibinfo {author} {\bibfnamefont{J.}~\bibnamefont{Klime{\v{s}}}}, \bibinfo
  {author} {\bibfnamefont{D.~R.}\ \bibnamefont{Bowler}},\ and\ \bibinfo
  {author} {\bibfnamefont{A.}~\bibnamefont{Michaelides}},\ }%
  \Doi{10.1088/0953-8984/22/2/022201}{\emph{\bibinfo {title} {Chemical accuracy
  for the van der Waals density functional}}},\ \bibinfo {journal} {Journal of
  Physics: Condensed Matter}\ \textbf{\bibinfo {volume} {22}},\ \bibinfo
  {pages} {022201} (\bibinfo {year} {2009}).~%
  \bibAnnoteFile{Stop}{Klimes2009}%
\end{thebibliography}
\end{document}